\documentclass[journal,comsoc]{IEEEtran}
\usepackage[T1]{fontenc}

\usepackage{xcolor}
\usepackage{fancyhdr}
\ifCLASSINFOpdf
  \usepackage[pdftex]{graphicx}
  \graphicspath{{./}{../pdf/}{../jpeg/}}
  \DeclareGraphicsExtensions{.pdf,.jpeg,.png}
  \usepackage[caption=false, font=footnotesize]{subfig}
\else
\fi

\usepackage{amsmath}
\usepackage[cmintegrals]{newtxmath}
\usepackage{algorithmicx}
\usepackage{algorithm}
\usepackage{algpseudocode}
\makeatletter
\newcommand{\removelatexerror}{\let\@latex@error\@gobble}
\makeatother

\newtheorem{definition}{Definition}

\begin{document}
\title{5G Enabled Fault Detection and Diagnostics: \\ How Do We Achieve Efficiency?}
\author{Peng~Hu,~\IEEEmembership{Member,~IEEE,}
        and~Jinhuan~Zhang\IEEEmembership{}
\thanks{P. Hu is with the Digital Technologies Research Center, National Research Council of Canada, Waterloo, ON N2L 3G1, Canada E-mail: Peng.Hu@nrc-cnrc.gc.ca}
\thanks{J. Zhang is with School of Information Science and Engineering, Central South University, Changsha 410083, China.}
}

\markboth{\copyright Crown copyright 2020. IEEE Internet of Things Journal, vol. 7, no. 4, pp. 3267-3281, April 2020, doi: 10.1109/JIOT.2020.2965034.}{  }


\maketitle

\begin{abstract}
The 5th-generation wireless networks (5G) technologies and mobile edge computing (MEC) provide great promises of enabling new capabilities for the industrial Internet of Things. However, the solutions enabled by the 5G ultra-reliable low-latency communication (URLLC) paradigm come with challenges, where URLLC alone does not necessarily guarantee the efficient execution of time-critical fault detection and diagnostics (FDD) applications. Based on the Tennessee Eastman Process model, we propose the concept of the communication-edge-computing (CEC) loop and a system model for evaluating the efficiency of FDD applications. We then formulate an optimization problem for achieving the defined CEC efficiency and discuss some typical solutions to the generic CEC-based FDD services, and propose a new uplink-based communication protocol called ``ReFlexUp''. From the performance analysis and numerical results, the proposed ReFlexUp protocol shows its effectiveness compared to the typical protocols such as Selective Repeat ARQ, HARQ, and ``Occupy CoW'' in terms of the key metrics such as latency, reliability, and efficiency. These results are further convinced from the mmWave-based simulations in a typical 5G MEC-based implementation.
\end{abstract}


%
\IEEEpeerreviewmaketitle

\section{Introduction}

\IEEEPARstart{W}{ith} the recent technological advancements in wireless communications and computing technologies, the 5th-generation wireless networks (5G), mobile edge computing (MEC), and Internet of Things (IoT) have become a driving force of a vast number of applications in industrial automation, oil \& gas, smart manufacturing, etc. Aligned with the vision of future Industry 4.0 and smart manufacturing systems, how to leverage 5G and MEC for industrial IoT (IIoT) has recently attracted intensive interests from industry and academia.

Of the particular interest to the IIoT is the use of wireless technologies for industrial automation, production, and control systems, which impose many challenges such as timing, reliability, and efficiency guarantees. For example, as a typical class of mission-critical industrial automation applications, real-time fault detection and diagnostics (FDD) systems \cite{Levrat2010, ISO2012} require timing and reliability guarantees for data collection, transmission, and processing. The data transmission latency usually needs to be maintained at the scale of milliseconds, which can hardly be met through classical wireless networking technologies. Employing wireless systems for industrial automation systems used in factory plants has been a challenge due to the complexity of factory settings, lack of best operational practice or IoT strategies, as well as security and performance concerns \cite{ONWorld2017}. Radio frequency identification (RFID) systems represent the classical wireless technologies used on information sensing devices \cite{Liu2017_cps} which have recently been applied in smart job-shops \cite{Ding2018}. Recent advancements in the low-power wide-area networks (LPWAN) \cite{Hoglund2017, IETF_lwpan_dt} and low-power wireless personal-area networks (LPWPAN) such as WirelessHART (i.e. IEC 62591) \cite{IEC2016}, ISA100.11a (e.g., IEC62734) \cite{IEC2012}, 6TiSCH \cite{6TiSCH2015, Palattella2015}, and Wi-SUN \cite{Wi-SUN2017} have shown the market-relevant advantages of applying the wireless technologies to industrial settings. However, many mission-critical applications demanding ultra-reliable low-latency communication (URLLC) \cite{ITU-R2015} cannot be addressed by these WPAN systems, as they cannot provide deep coverage, scale-up deployments, and efficient operations for scalable and geographically distributed factory plants. Therefore, alternative solutions need to be secured. Fortunately, 5G URLLC aims to meet this demand although it requires new designs. The URLLC schemes based on coding \cite{Wu2018, Sybis2016}, relaying \cite{Swamy2017, Hu2018}, multiple access\cite{Ma2018, Shariatmadari2016, Tian2017}, and interface/channel/user diversity \cite{Singh2018, Nielsen2018, Jurdi2018} have been proposed in the literature. URLLC-based control schemes are discussed in \cite{Voigtlander2017, Jurdi2018, Liu2018} and its transmission and resource optimization strategies are introduced in  \cite{Shariatmadari2016}. 

However, URLLC alone does not necessarily guarantee the efficient execution of mission-critical applications. From the network architectural perspective, MEC can help a 5G URLLC system achieve application-specific efficiency, with which low latency and high reliability in communication and efficient computation can be managed. In addition, MEC can help achieve the latency performance that is hardly made with a typical cloud computing based IIoT architecture \cite{Hu2015}. Putting the communication to the edge will enable the fast communication with the local cloudlets and endpoints. Data analytics, device management, automation, and artificial intelligence (AI) are the typical IoT Edge use cases reported in \cite{GSMAIoT}, where protection of data privacy and security risks such as denial-of-service (DoS) attacks and authentication can be better handled with the closer monitoring of the data flows at the edge.

\begin{figure}[!t]
\centering
\includegraphics[width=3.6in]{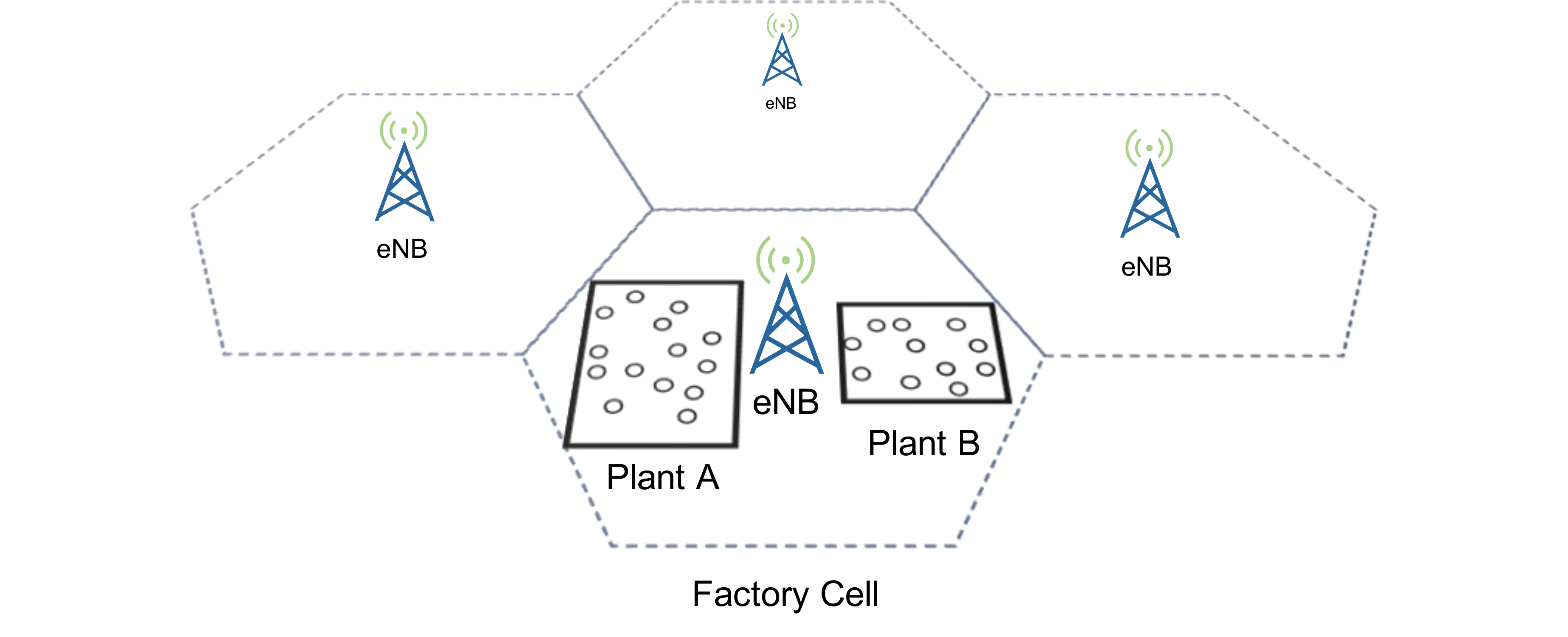}
\caption{Example of a factory cell outlined by the dotted lines as coverage area of an evolved Node B (eNB) node, where each cell may have multiple plants denoted by square blocks within which sensors or actuators denoted by circular nodes are deployed.}
\label{Fig1}
\end{figure}

Although the 5G URLLC paradigm and MEC shed light on solving the challenges of many mission-critical industrial automation systems, the fundamental question ``how can we achieve efficiency of FDD-based industrial applications enabled by 5G URLLC and MEC?'' is still left unanswered. This paper aims to answer this question in several aspects with the following contributions:

\begin{itemize}
\item We have explored the efficient execution of an FDD service (FDDS) as a representative class of 5G-enabled IIoT applications based on an MEC architecture.

\item We have proposed the concept of the communication-edge-computing (CEC) loop to incorporate the factors of the URLLC-based data transmissions and the statistical data analysis at the mobile edge. The proposed CEC loop provides a framework for analyzing IIoT systems based on URLLC and MEC.

\item We have formulated an optimization problem of achieving the CEC efficiency and discussed the solutions in some typical cases in the context of FDDS.

\item We have proposed a protocol for the reliable and flexible uplink (UL) communication called ``ReFlexUp'' that can be used for the URLLC-based IIoT systems where UL traffic is dominant. The ReFelxUp utilizes the analytical results of the proposed CEC efficiency and its performance of ReFlexUp is validated with numerical and simulation results and compared against the classical schemes, such as Selective Repeat automatic repeat request (ARQ), Hybrid ARQ (HARQ), and Occupy CoW \cite{Swamy2017}.

\end{itemize}

The rest of the paper is structured as follows. Section II presents the related work. Section III discusses the FDD problem in the industrial settings. Section IV proposes a system model for 5G MEC-based FDD services. Section V presents the proposed communication protocol. Section VI gives the performance analysis of the classical protocols. Section VII discusses the evaluation results. Conclusive remarks are made in Section VIII.


\section{Related Work}

The requirements for the real-time capability on the integrated computation and communication systems have been seen in the cyber-physical systems \cite{Liu2017_cps} and factory automation systems. Classical industrial automation systems use wired communications, such as the Fieldbus, Highway Addressable Remote Transducer Protocol (HART), and industrial internet technologies. Wired medium provides a reliable connection for various mission-critical tasks in a field network. Using wireless systems in the industrial settings can provide several advantages including reducing the amount of time and efforts on installation and deployment; however, traditional wireless systems cannot be suited for such tasks mainly due to the concerns such as reliability, performance, and the lack of the best operational practice.

In the recent decade, IoT has gained much attention in the industry \cite{Wi-SUN2017}, and new installations of wireless networks for industrial applications have been increasing \cite{IEC2014WP, ONWorld2017}. With the introduction of WirelessHART \cite{IEC2016} and ISA100.11a \cite{IEC2012} protocols, employing LPWPAN systems has become a popular option for regular  manufacturing automation systems. WirelessHART and ISA100.11a aim to provide reliable end-to-end data transmissions with the time slotted channel hopping (TSCH) based schemes and they both have been broadly used in the industrial automation systems. The IEEE 802.15.4e specification \cite{IEEEStd802154} then adds a MAC-layer enhancement with the TSCH scheme to provide new features for industrial applications. Based on the MAC-layer enhancement in IEEE 802.15.4e, IETF 6lo working group ``6tisch'' has proposed solutions \cite{6TiSCH2015, Palattella2015} to support IPv6 over low-power WPAN (6LoWPAN). In general, an LPWPAN-based system provides data link-layer reliability and network-layer redundancy with mesh networking and has been proven effective in industrial automation systems. Based on IEEE 802.15.4g, Wi-SUN \cite{Wi-SUN2017} arises to provide solutions to smart utility networks and other IIoT applications. A smart job-shop system based on RFID tags \cite{Ding2018} is presented, where heterogeneous production data mining tasks from various RFID event data have been discussed.

LPWANs provide another approach to the IIoT systems. LPWAN is a generic name of a new set of wireless networking technologies which can enable massive machine-type communication (mMTC). Two typical standardized LPWAN technologies are 3GPP Long Term Evolution category M1 (LTE-M) and narrowband IoT (NB-IoT) \cite{Hoglund2017}. The IETF working group ``lpwan'' \cite{IETF_lwpan_dt} has been formed to make efforts on network-layer solutions to secure operations and device management. LTE-M and NB-IoT provide options for various industrial applications requiring different data rates, latency, and communication modes using low-cost user equipment (UE).

However, currently the aforementioned LPWPAN and LPWAN technologies can hardly be applied in the mission-critical systems requiring very low latency, extended communication coverage, and support for various communication scenarios. For example, WirelessHART can hardly reduce the latency to less than 10 ms. This challenge has been noticed aligned with most recent advancement under the umbrella of 5G. ITU-T \cite{ITU-R2015} and 3GPP \cite{Carugi} have made 5G move toward three distinctive paradigms, i.e., enhanced Mobile Broadband (eMBB), mMTC, and URLLC, where URLLC aims to meet the requirements of many mission-critical industrial automation systems and tactile Internet with the overall target of 99.999\% reliability and 1 ms latency. 

There are many technical challenges in URLLC to overcome including the new theories and designs from the physical layer in UL or downlink (DL) communications especially for 5G-enabled IIoT systems working in a complex radio environment. Reliable data transfers can be achieved through the error mitigation with coding, retransmission, diversity, multiuser, and cooperative communications. A coded tandem spreading multiple access (CTSMA) is proposed in \cite{Ma2018}. In the UL communication, a collision-reduction scheme based on transmitting same packets across multiple channels is proposed in \cite{Singh2018}. An optimal transmission and resource allocation based on ARQ and HARQ for ultra-reliable communication is proposed in \cite{Shariatmadari2016}. With the presence of multiple network interfaces, a multi-interface diversity scheme is proposed \cite{Nielsen2018} to distribute coded payload over multiple interfaces. Relay protocols provide a different way of meeting URLLC requirements. A two-phase control scheme for DL in the factory automation systems is proposed in \cite{Jurdi2018, Liu2018} in order to achieve the URLLC requirements. Authors of \cite{Swamy2017} devise a new protocol called ``Occupy CoW'' to achieve the 99.999\% reliability of UL/DL data transmissions and low latency with the wireless networks. Although the two-phase relaying protocol can meet the latency and reliability requirements, it can hardly address the 5G or MEC implementations in typical sensing scenarios for an industrial automation system which requires a combination of sensor devices, including small sensors and actuators powered by batteries or energy harvesters.

5G-based MEC can provide a fundamental architectural support for the mission-critical applications such as predictive maintenance and monitoring for numerous industrial processes. MEC allows the local processing of data at the mobile edge of 5G networks within one or more cells covering factory plants as illustrated in Fig. \ref{Fig1}, where each cell can have multiple plants. The ETSI MEC GS 003 \cite{ETSI2016} specifies a reference architecture where various mobile edge apps on top of a virtualization infrastructure on a mobile edge host. A proof-of-concept MEC system (mecwiki.etsi.org) is promised to enable the trusted and real-time data delivery between the IoT devices and cloud-based applications. The ETSI-compliant design of an MEC platform based on containers/VMs and Open vSwitch (OVS) solution is proposed in \cite{Hsieh2018}, where the MEC platforms are interconnected through OVS devices. The MEC architecture designs have been recently discussed in the literature. Authors of \cite{KaiVictorLeung2018} discuss an MEC design where the MEC server sits between mobile users and the core network. The GSM Association (GSMA) \cite{GSMAIoT} indicates the three options of the IoT edge locations from the mobile operator's perspective, which, listed by the increasing distance to an IoT device, are on the edge server/node/gateway, on the base station (BS), or on the distributed local data center between the BS and the network core. In \cite{Tong2016}, a hierarchical edge cloud architecture consisting of 3 tiers of edge cloud servers is proposed, where the tier-1 server is close to the users through the wireless links. In \cite{Cao2019},  the edge servers are considered to be on the 5G BS nodes which connect end devices such as cameras. In \cite{Ford2017}, the mobile edge cloud is placed to the public data network (PDN) side connected to the 5G core network.

The aforementioned architecture designs of MEC follow a similar pattern that an MEC platform is deployed close to the UEs on the cellular networks. We adopt this pattern in the paper: an MEC node is put in proximity of a cellular BS (i.e., eNB) as a standalone entity which connects to the eNB through a high-bandwidth point-to-point (P2P) link with extremely low latency. This is close to a practical setup from the mobile operator's perspective in that it remains minimum changes and backward compatibility to the existing radio access network (RAN) infrastructure and it is flexible enough to adjust the deployment parameters (e.g., distance and link bandwidth) between the MEC node and the eNB.

URLLC can provide communication service required by an FDDS but it alone does not necessarily ensure an efficient FDDS process. This is because: (1) an FDDS needs to be made with accurate and sufficient data within a time frame; (2) multiple FDDS tasks need to be considered; and (3) a communication process needs to adapt to an FDDS process at the mobile edge.

The previous works have addressed some important aspects surrounding 5G URLLC systems, but additional application-specific requirements need to be met with not only the data transfers based on URLLC but also the computation tasks running at the mobile edge. For example, as a typical type of 5G-enabled mission-critical applications in industrial automation, fault detection and diagnostics (FDD) systems require real-time performance, accuracy, and efficiency, which are determined by the wireless communication between sensor nodes and MEC servers (or platforms) as well as the FDD computation process running at the MEC server. Therefore, we need to address this challenge considering the 5G communications and the MEC for FDD solutions. If we consider the MEC-based FDDS within the factory plant as a control loop, we need to fundamentally understand the optimal strategy regarding the URLLC-based data transfers and the FDDS.

To the best of our knowledge, this paper first proposes a framework and optimal strategies for URLLC data transfers and edge computation tasks. Based on the framework, we will propose a protocol that address a typical FDD scenario in industrial automation.

\section{Fault Detection and Diagnostics for Industrial Automation}
FDD is aligned with the ISO 13374 standards \cite{ISO2012} which specify the information reference architecture for condition monitoring and diagnostics (CM\&D) applications and the proposed open system architecture in \cite{Levrat2010}, consisting of data acquisition, data manipulation, state detection, health assessment, prognostics, decision support and presentation. A typical FDD system is implemented through a central entity where sensors are installed into a distributed control system (DCS) to report their data periodically to the central entity running a statistical FDD scheme. With the 5G-based MEC, a new architecture can be devised where we can move such a central entity close to the BS of a mobile network, e.g., eNB in Fig. \ref{Fig1}, and an MEC server handles the sensing data for an FDDS task and executes the computation jobs for the task. The computation jobs use the allocated compute resources at the MEC server for an FDDS task.

\begin{figure}[!t]
\centering
\includegraphics[width=2.8in]{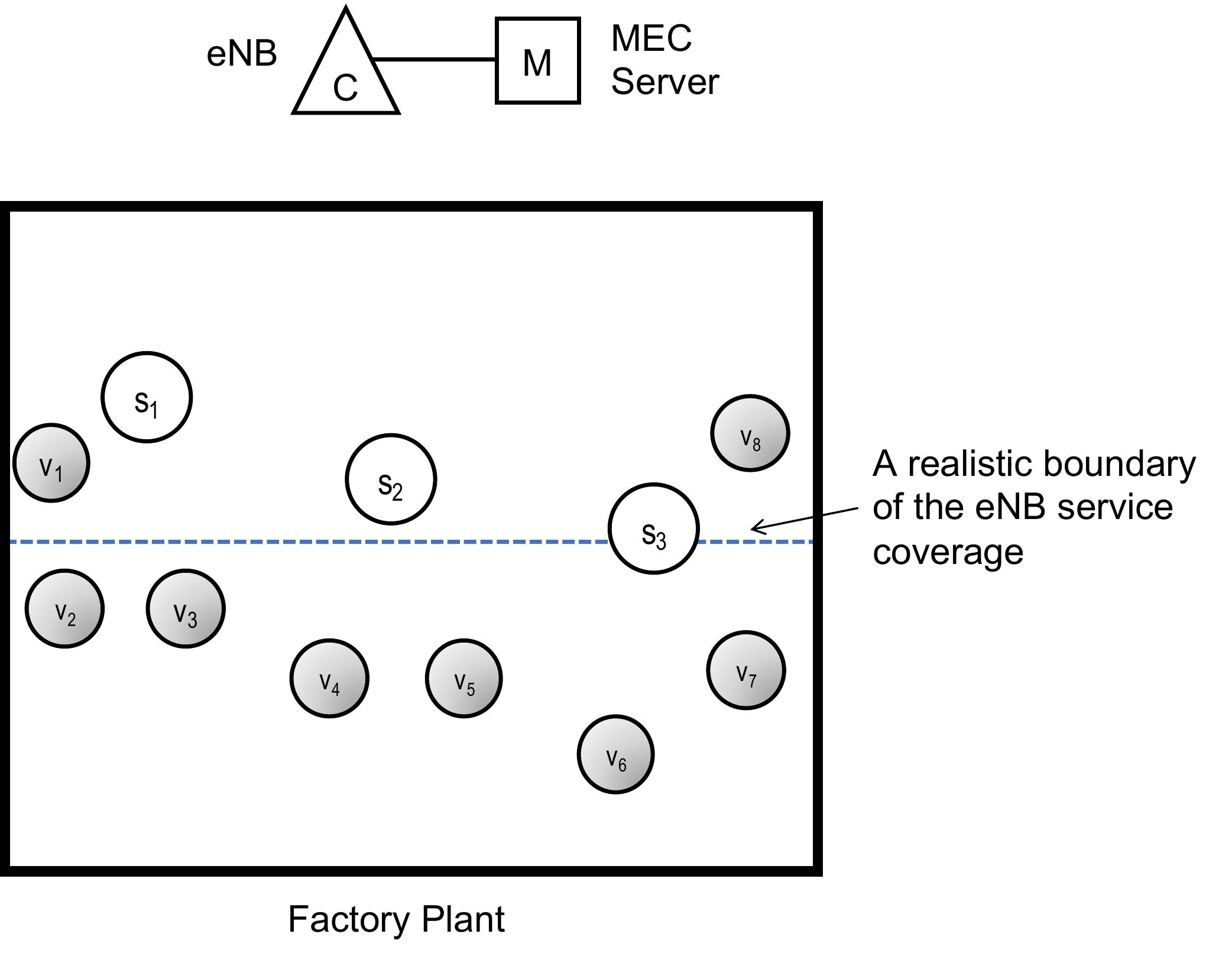}
\caption{Illustration of the network with an eNB (labeled as C) and an MEC server (labeled as M), three relay nodes (labeled from $s_1$ to $s_3$), and eight sensor nodes (labeled from $v_1$ to $v_8$) deployed in a factory plant.}
\label{Fig2}
\end{figure}

\begin{table}[H]
\caption{Notation Definitions}
\label{Tbl:NotationDef}
\centering
\begin{tabular}{l | l}
\hline
\textbf{Notation} & \textbf{Definition}\\ 
\hline
ARQ & Automatic repeat request\\
BS & Base station\\
CEC & Communication-edge-computing\\
CTSMA & Coded tandem spreading multiple access\\
CM\&D & Condition monitoring and diagnostics\\
DCS & Distributed control system \\
DoS & Denial-of-service\\
DL & Downlink\\
eMBB & Enhanced mobile broadband\\
eNB & Evolved Node B\\
FDD & Fault detection and diagnostics\\
FDDS & FDD service\\
EPC & Evolved packet core\\
GSMA & GSM Association\\
HART & Highway Addressable Remote Transducer Protocol\\
IIoT  & Industrial Internet of Things\\
LPWAN & Low-power wide-area networks\\
LTE & Long term evolution\\
MEC & Mobile edge computing\\
mMTC & Massive machine-type communication\\
MCL & Maximum coupling loss\\
OSI & Open Systems Interconnection\\
OVS & Open vSwitch\\
P2P & Point-to-point\\
PCA & Primary component analysis\\
PDN & Public data network\\
PV & Process variable\\
RAN & Radio access network\\
RB & Resource block\\
ReFlexUp & Reliable and flexible uplink communication\\
RFID & Radio frequency identification\\
RLC & Radio link control \\
SNR & Signal-to-noise ratio\\
SPE & Squared prediction error\\
S/P-GW & Serving/packet gateway \\
TSCH & Time slotted channel hopping\\
UE & User equipment\\
UL & Uplink\\
URLLC &  Ultra-reliable low-latency communication\\
VM & Virtual machine\\
VP & Virtual processor\\
WPAN & Wireless personal-area networks\\
6LoWPAN &IPv6 over WPAN\\
LPWPAN & Low-power WPAN\\
\end{tabular}
\end{table}

\subsection{An MEC-based FDDS Architecture}

An MEC-based FDDS architecture is depicted in Fig. \ref{Fig2}, where an FDDS is deployed at the mobile edge at the MEC server, which includes a mobile edge host \cite{ETSI2016} attached to the eNB as a central controller C, and a local factory cell consisting relay or controller nodes, and sensor nodes. In a real-world scenario, it is possible that some sensor nodes may be beyond the coverage of C which will result in the failed delivery of critical sensing data required by the FDDS. This case is illustrated in Fig. \ref{Fig2}, where $v_2$ to $v_7$ is beyond the service coverage of C due to the maximum coupling loss (MCL) boundary shown in dotted lines. As a general method, the use of relay nodes can alleviate such problem as proven effective in \cite{Swamy2017}. 

\begin{figure}[!t]
\centering
\includegraphics[width=3.4in]{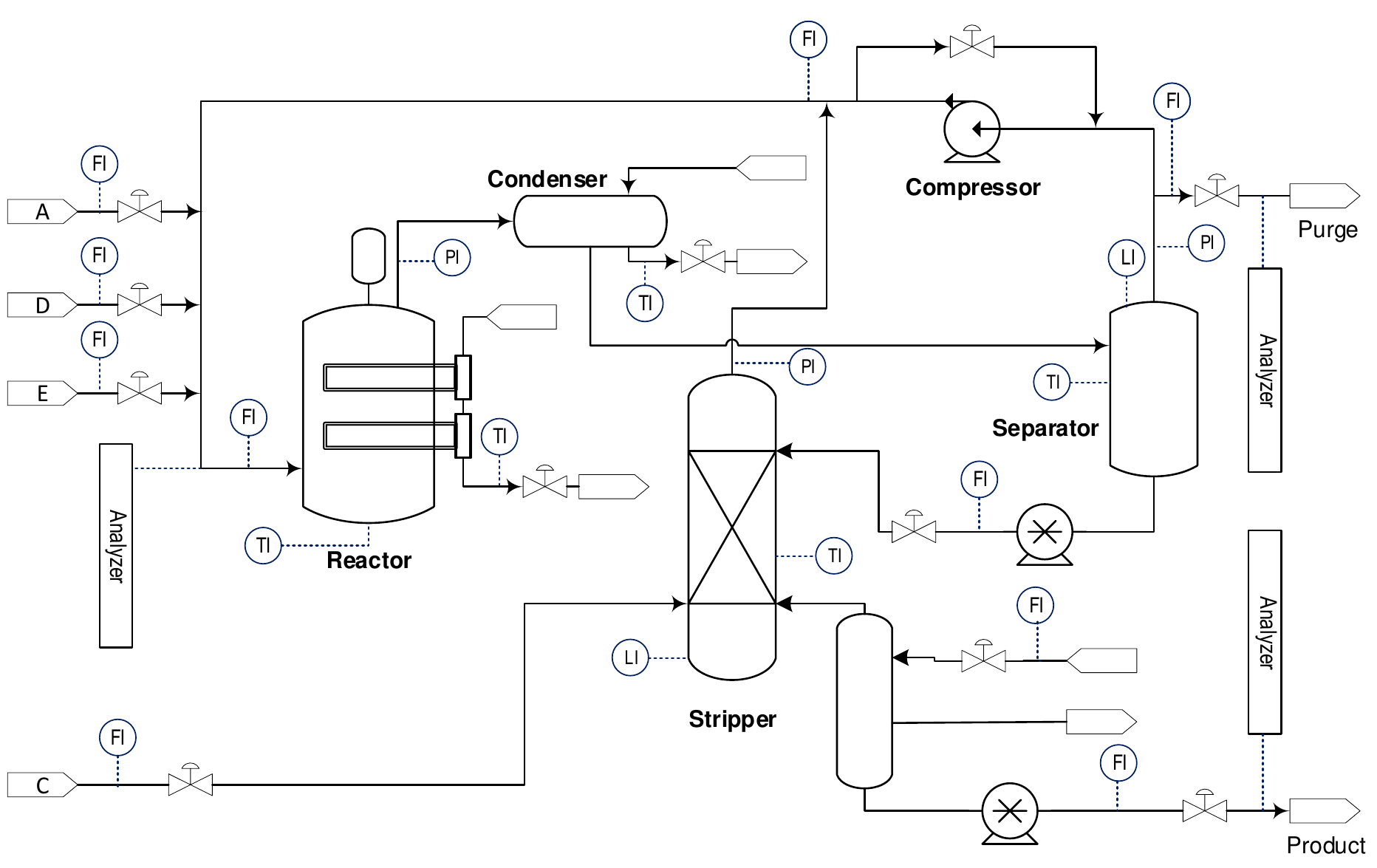}
\caption{An example real-world sensor system deployment based on the TE process. }
\label{FigTEP}
\end{figure}

Let us look at a real-world deployment of FDDS in a broadly used benchmark Tennessee Eastman (TE) process model \cite{Downs1993} as shown in Fig. \ref{FigTEP}, which is a broadly used model for benchmarking industrial processes, monitoring, and operations. In Fig. \ref{FigTEP}, there are four inputs, A, D, E, and C, and five main components for producing chemical products, i.e., reactor,  condenser, stripper, separator, and compressor, where the product is shown at the bottom right output symbol and sensors can be installed onto the process control line to monitor the key process variables (PVs). A typical TE process model \cite{Downs1993} has 41 measurements and 12 manipulated variables, where these 53 measurement/manipulated PVs can be obtained from the sensors, installed in the interfaces and analyzers that are connected to the TE process system shown in dotted lines shown in Fig. \ref{FigTEP} and other process lines. Each sensor is deployed to generate a time-series data stream or flow for one type of PVs at intervals. In addition, a sensor can be extended to generate different types of PVs at different intervals, which over time results in generation of multiple data flows from the sensor. 

Aligned with Fig. \ref{Fig2}, the sensing data is received at the mobile edge for an FDDS where there are $\mathcal{N} \geq 1$ FDDS tasks up and running and each task deals with a stream of sensing data.

\subsection{Computation Jobs for FDDS Tasks}
\begin{figure}[!t]
\centering
\includegraphics[width=3.4in,height=2.1in]{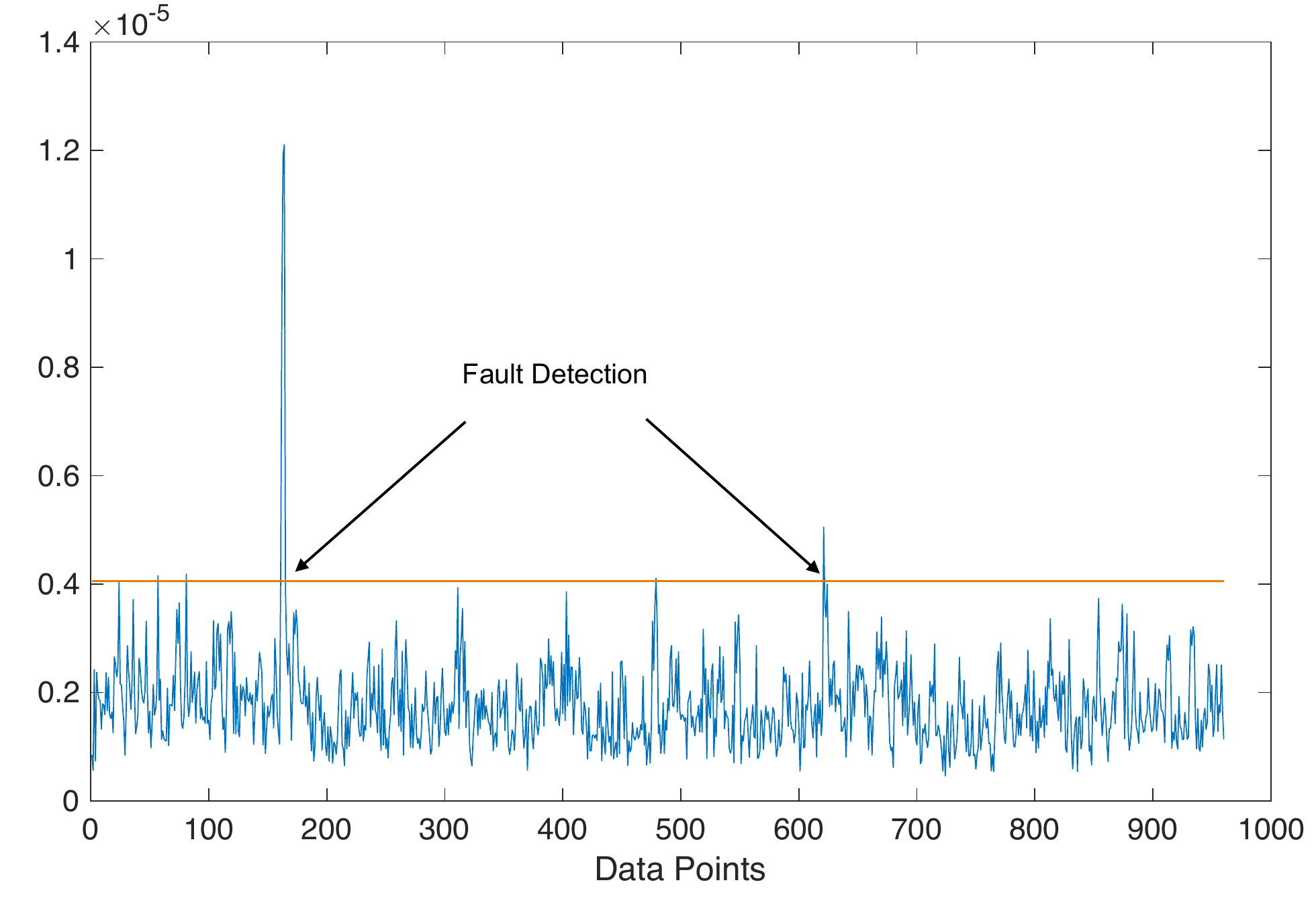}
\caption{Example result based on the Squared Prediction Error (SPE) for fault detection and diagnostics (FDD) from the sample data of a typical TE process.}
\label{FigTepFdDemo}
\end{figure}

Balancing the computation load and communication latency is an important challenge in MEC \cite{Mach2017}. Here let us see how a real-time FDDS is performed based on the standard principle component analysis (PCA) \cite{Yin2012}, where two steps are involved: off-line analysis and on-line analysis. The first step requires us to do the standard PCA out of the normal data samples with no faults we collected through a URLLC process and derive the values of the squared prediction error ($\text{SPE}_0$) and Hotelling's T-Squared ($T^2_0$). The second step requires us to do the on-line collection of new data samples in real time and to calculate the new $\text{SPE}(t)$ and $T^2(t)$ at the time $t$. If $\text{SPE}(t)$ or $T^2(t)$ is greater than $\text{SPE}_0$ or $T^2_0$, an error is detected. For example, Fig. \ref{FigTepFdDemo} shows the result of a fault detection in the on-line analysis based on the model established in the first step, using the broadly-used TE process data set (http://web.mit.edu/braatzgroup) where the data samples of normal and erroneous operations are given. Based on the TE process and sensors deployed in the process line, then fault diagnosis can be conducted at an MEC platform with the knowledge database and the possible sensor locations of the occurrence. Therefore, fault detection is essential for a successful FDD process and for the FDDS deployed at the mobile edge.

Based on the data samples, if we let the number of components be 17 and the significance level be 0.01 (which corresponds to three-sigma rule of thumb), the result is shown in Fig. \ref{FigTepFdDemo}. We can see that an accurate and effective FDD relies on the sufficient data samples when executing FDDS tasks. Thus, an effective FDDS task needs to consider not only the transfer of data samples in packets through a URLLC process, but also the mechanism to guarantee the required in-order data samples to make successful computation job done within a time period $T_p$ where the FDDS is done on a time-sharing basis. In addition, we need to consider a general case where multiple FDDS tasks are executed during the time period $T_p$. Therefore, there is a need for looking for an optimum solution to the FDDS tasks. We will discuss this typical industrial application in more detail starting with a system model. 

\section{System Model}

\begin{figure}[!t]
\centering
\includegraphics[width=3.6in]{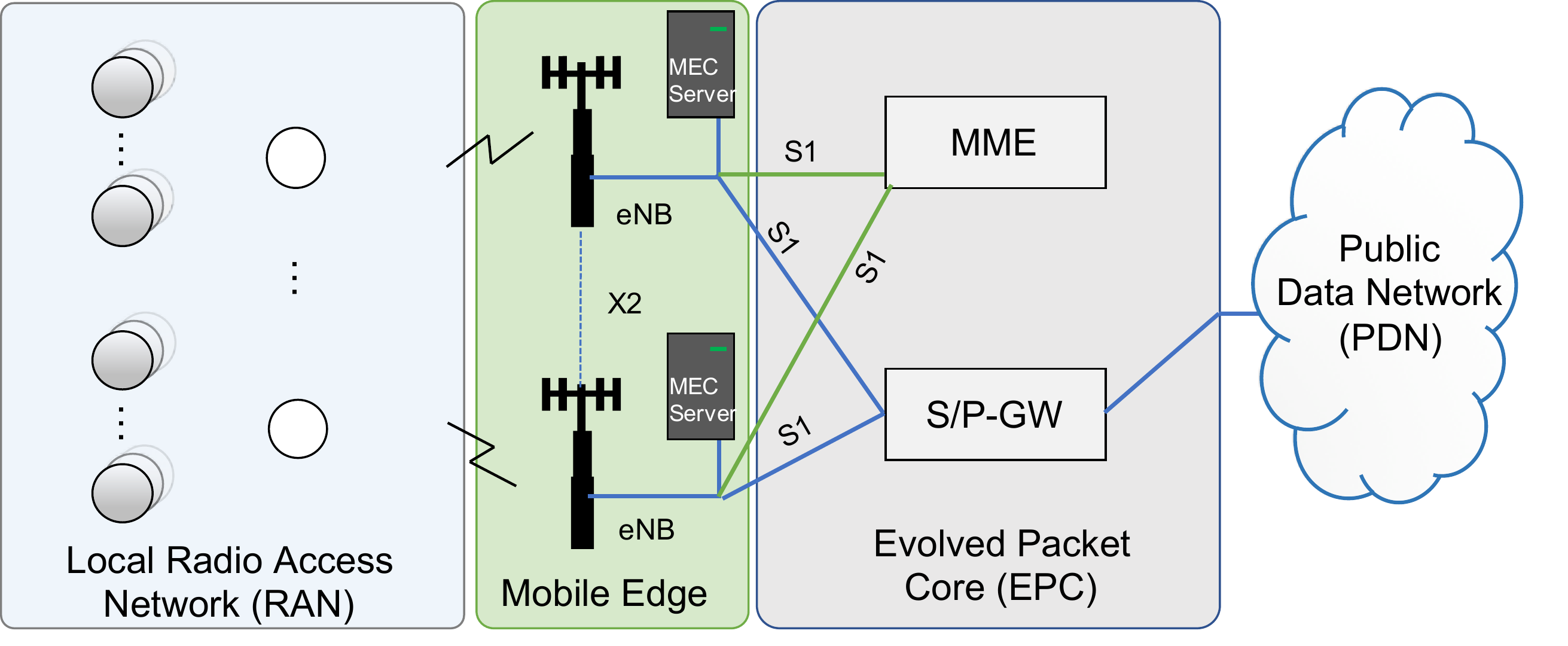}
\caption{Example deployment of a proposed 5G MEC based system, where two MEC server nodes are co-located with two 5G eNB nodes connected to the EPC entities, and the relay/sensor nodes are in a RAN.}
\label{Fig5GMecDeploy}
\end{figure}

A system deployment for 5G-based factory automation system is shown in Fig. \ref{Fig5GMecDeploy}, where we can have 5G-compliant sensors and actuators as relay and sensor nodes deployed on the RAN located on the shop floor for monitoring/diagnostic purposes, and a mobile edge entity consisting of a few 5G eNB nodes with MEC servers. The MEC servers can be connected to the evolved packet core (EPC) via the S1 interface including the serving/packet gateway (S/P-GW) that can transfer the data-plane traffic to a public data network (PDN). In addition, the data traffic characteristics on the RAN based on Fig. \ref{Fig5GMecDeploy} is UL dominant, which indicates that the amount of sensing data to be transferred from a UE node to its eNB via the UL is greater than that of the control data via the DL.

Aligned with the deployment in Fig. \ref{Fig5GMecDeploy}, we use the network shown in Fig. \ref{Fig2} in the analysis here, where we denote the entire field wireless network by $G$, where $G$ contains two sets of nodes: one is $S$ (i.e., the set of relay nodes) and the other is $V$ (i.e., the set of sensor nodes), $G=S \bigcup V$ and $ S \bigcap V = \emptyset$; each node in $S$ has a radio interface for communicating with the associated eNB denoted by C, which is interconnected to an MEC platform M; and the $i$-th relay node can function as a local controller for its member sensor nodes in $V_i$, $i \in \{1, ..., |S|\}$. The total number of nodes in $G$, $S$, and $V$ are denoted by $N_G$, $n_s$, and $n_v$, respectively.

In reality, at least two-phase communication is needed for the indoor communication as shown in Fig. \ref{Fig1}. As illustrated in Fig. \ref{Fig2}, if the coupling loss (i.e., the long-term channel loss over a link including path loss, shadowing, etc.) in the DL between C and nodes in $V$ is the same, $v_2$ to $v_6$ may be beyond the service coverage of C (i.e., eNB) without the relaying effort of $s_1$, $s_2$, or $s_3$; and, in fact, that coverage at different locations of a faculty cell varies due to the complex environment. The current 3GPP NB-IoT can have a coupling loss as much as 164 dB \cite{Hoglund2017}, meaning a node experiencing coupling loss less than 164 dB can be considered in a cell. The coupling loss in the indoor radio environments as shown Fig. \ref{Fig1}, from ITU-R P.1238-9 \cite{ITU2017} the path loss affected the indoor power loss and penetration factors may need to be compensated by the transmit power. However, the transmit power on sensors, especially for battery powered sensors, is usually not feasible; or, if the power is increased, the interference to other nodes may be significant. The indoor factory environment also has complex fading conditions that would cause a link failure on one or more nodes, although they seemly have connections to C. Similarly, the aforementioned circumstances can apply to the UL between C and nodes in $V$.

\subsection{The CEC Loop}

\begin{figure}[!t]
\centering
\includegraphics[width=3.2in]{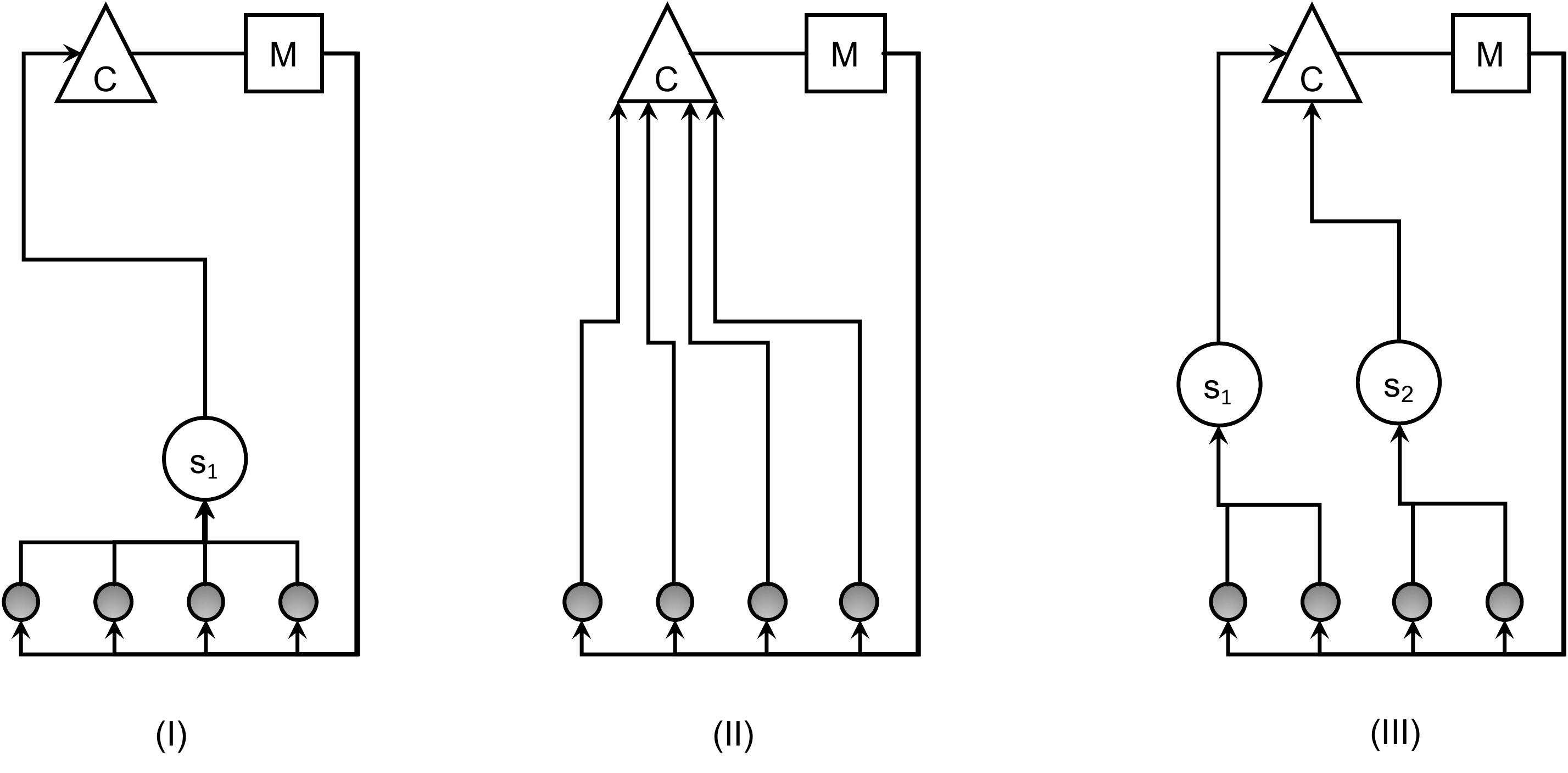}
\caption{Three basic models of a CEC loop: (I) shows sensor nodes and C are connected indirectly with a replay node; (II) shows sensors and C are connected directly; and (III) shows two relay nodes connect sensor nodes in two sets. }
\label{FigFModels}
\end{figure}

To abstract the possible models of the MEC-based computing with 5G, we propose the concept of the CEC loop. From the data flow perspective, the concept of a CEC loop can be shown in the three models in Fig. \ref{FigFModels}, where the model (I) in Fig. \ref{FigFModels}(I) shows that all sensing data flows to the relay node $s_1$ and then to C and M, and then the feedback information from C will be sent to $v$ nodes in $s_1$; the model (II) in Fig. \ref{FigFModels}(II) shows the typical data flow in a star topology, where there are no relay nodes between C and $V$; and the model (III) in Fig. \ref{FigFModels}(III) shows an extended case of Fig. \ref{FigFModels}(I) where nodes in $V$ are split into two sets, handled by two relay nodes, $s_1$ and $s_2$, which can shift the loads of $s_1$ in Fig. \ref{FigFModels}(I).

In this paper, we adopt the model (III) because: (a) although three models require the active receiver (RX) mode on sensor nodes,  nodes in $V$ conserve power in the transmitter (TX) mode compared to model (II); (b) the model (III) can be set to backward- and forward-compatible with legacy systems (e.g., existing Fieldbus networks) and new systems, respectively; (c) the use of relay nodes can ease the system installation and integration efforts; (d) $s$ can be equipped with multiple antennas and addition power to ensure the available links between $s_i$ and $v_i$ nodes and between $s_i$ and C; (e) the model (III) is generic enough which can be considered the distributed version of (I); and (f) there are additional performance advantages of the model (III), including the performance improvements, such as mitigation of jitters and distributed local decisions.

In a real-world industrial network following the model (III), the number of relay nodes deployed in the field is adjustable according to the sensor nodes in $V$, which can be added dynamically in a process or machine monitoring system.

\begin{definition}[CEC Loop Time] The total time period including the time taken to transmit a required stream of data packets to M through a central controller C at the mobile edge, and the computation time taken to process the data stream.
\end{definition}

Based on the definition, if we denote the CEC loop time by $T_p$, it can be expressed in (\ref{eqn:Tp}), which implies three parts: (1) the time that a stream of data takes to be delivered to M, i.e., $T_{cm}$, and (2) the time taken at M to process the data received for an FDD process, denoted by $T_{cp}$, and (3) after the FDD process, the time $T_{fb}$ taken to deliver feedback messages to the local sensor or controller nodes. Since $T_{fb}$ is either an optional or small value, it is considered not a dominant factor and it will not be included in our discussion. In addition, to focus on our discussion on the major CEC factors, the data acquisition time of sensor nodes and the queuing/buffering delay on relaying nodes will not be not discussed. However, from an initial analysis based on the Little's theorem, we know the average queuing delay is mainly dependent on the queue size and packet arrival rate at the relay node, and therefore we can assume it is a small constant at a relay node.

\begin{equation}\label{eqn:Tp}
T_p = T_{cp} + T_{cm} + T_{fb}
\end{equation}

\begin{definition}[Reliability of the CEC Loop] For the $i$-th task, the ratio of the received number of packets $d(i)$ to the number of required packets $D(i)$ within the time period of $T_p$ needs to be met at a certain level. 
\end{definition}

If we denote the reliability of the CEC by $\varepsilon$,  $0 \leq \varepsilon \leq 1$. Although we expect to achieve  $\varepsilon = 1$ to have a full stream of data for analysis, for some applications, it might be valid if $\varepsilon < 1$.

\begin{definition}[Communication Failure] The event that one or more nodes in $G$ fail to deliver data to M during $T_p$ given a value of $\varepsilon$. The probability of such communication failure is denoted by $P_{cf}$.
\end{definition}

\begin{definition}[FDDS Task Failure] The event that at least one of the packets required for stream data processing is missing with the time period $\varepsilon$. The probability of such communication failure is denoted by $P_{df}$.
\end{definition}

Based on the aforementioned two definitions, the CEC failure can be calculated by $P_{cf} \cdot P_{df}$.

\subsection{CEC Efficiency}

In order to achieve an efficient FDDS process in the CEC loop, first we need to measure the performance of FDDS tasks in terms of the utilization of the compute resource (e.g., a virtual processor on M) and the communication resources (e.g., a resource blocks). Then, we need to derive a system model that can capture the task executions at M. Next, a problem needs to be formulated for its application into a generalized FDDS-like scenario.

\begin{figure}[!t]
\centering
\includegraphics[width=3.0in]{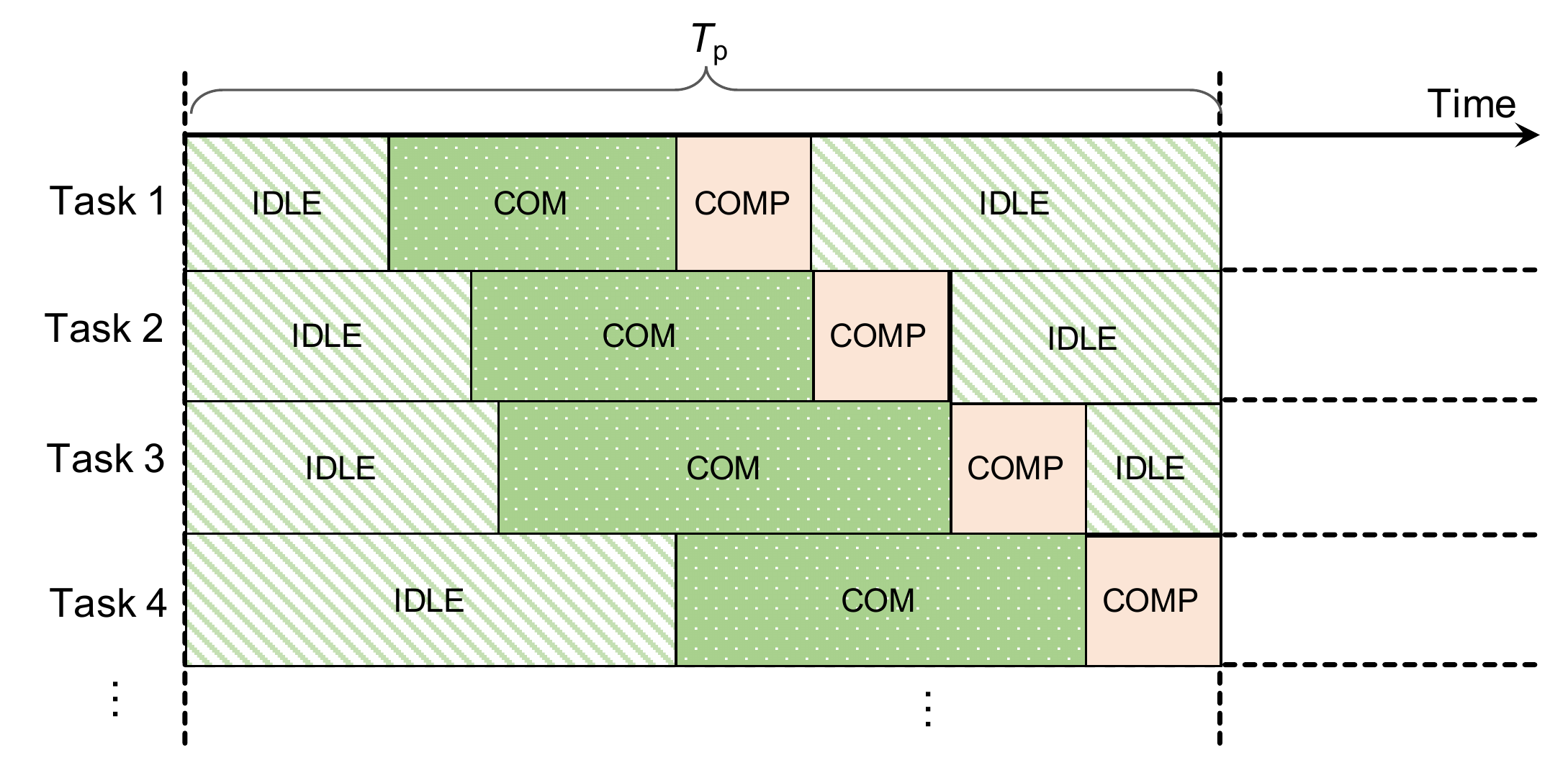}
\caption{Illustration of the task execution on a virtual processor at M, where the `COM' and `COMP' blocks represent the time periods for communication of a data transmission process and for computation of a task execution process, respectively. }
\label{FigCommComp}
\end{figure}

The definition of the CEC efficiency is given as follows:

\begin{definition}[CEC Efficiency] is the product of the virtual processor utilization used by all $\mathcal{N}$ tasks running at M and the utilization of $K$ resource blocks (RBs) used by these tasks.
\end{definition}

\begin{equation}\label{eqn:uc_definition}
u_c(i)=\dfrac{T_{cp}(i)}{ T_{cp}(i) + T_{cm}(i)}
\end{equation}

If we let $u_{c}(i)$ be the compute resource utilization of the $i$-th task at M defined in (\ref{eqn:uc_definition}), where, within a time slot $T_p$, $T_{cp}(i)$ is the time spent on computation jobs (e.g., data analytics) for the $i$-th task, and $T_{cm}(i)$ is the time spent on the required communication process for task $i$ in order to execute its FDDS task. This definition reveals the fact of an important type of generic industrial applications: for a computation task that takes some processing time at M, there requires a procedure of preparing the data which is obtained through a URLLC process. 

From (\ref{eqn:uc_definition}), we can derive some useful properties with their physical meanings. First, we can see that $u_c(i)$ is inversely proportional to $T_{cm}(i)$, meaning that the longer it takes to obtain data, the lesser the compute resource at M can be utilized. Second, we can see that when $T_{cm}(i)$ has a large enough value, $u_c(i)$ approximates to zero. It makes sense because if the required data needs a very long time to be delivered to M, M gets almost no chance to use a compute resource to process it. However, for real-world applications, we need to consider the facts: (a) $T_{cm}(i)$ has a boundary $T_p$; (b) there can be a timeout strategy we often use at M, i.e., M sets up a timer so when $T_{cm}(i)$ times out, an event will be triggered at M to solicit the missing data via a communication process; (c) $T_{cp}(i) = 0$ and $u_c(i)=0$ when there is no communication, i.e., $T_{cm}(i) = 0$; and (d) when the communication process takes very little time, $u_c(i)$ approximates to 100\%, i.e., $\lim_{T_{cm}(i) \to 0} u_c(i)= 1$. Therefore, we can see that in reality $0 < T_{cm}(i) < T_p$, $0<u_c(i)<1$. When $T_{cm}(i)=0$ or $T_{cm}(i)=T_p$, $u_c(i)=0$. In this sense, (\ref{eqn:uc_definition}) can well reflect the cases including boundary conditions in reality.

If we let $u_{RB}(i)$ be the RB utilization of the $i$-th task, then the CEC efficiency $U_{cc}$ is expressed as follows:

\begin{equation}\label{eqn:ucc}
U_{cc}=\sum_{i=1}^{\mathcal{N}} u_{c}(i) \cdot  u_{RB}(i)
\end{equation}

In (\ref{eqn:ucc}), let us see why $U_{cc}$ can measure the efficiency of a communication and computation process. Suppose the total number of tasks per time slot $T_p$ is $\mathcal{N}$, and all these tasks use a processor unit which is considered as a compute resource (note here we neglect the memory usage on M as it is usually sufficiently large). $u_{RB}(i)$ captures the important communication resource used by an FDDS task, of which an example execution is shown in Fig. \ref{FigCommComp}.

In order to consider the relationship between the $i$-th task and its RBs used, we define the set of the RBs used by the $i$-th task as: $k(i) \subseteq \mathbf{K}, ~\mathbf{K} = \{1, ..., K\} $, for $i \in \mathbf{N},~\mathbf{N} = \{1, ..., \mathcal{N}\}$. In a general case, one task can take a portion of one or more RBs, therefore $|k(i)| = \sum_{j=1}^{K} I_{RB}(i,j) $, and $0 \leq |k(i)| \leq K$, where $I_{RB}(i,j)$ is the indicator function defined in (\ref{eqn:NRB_indicator_func}) and it is dependent on an RB scheduling policy.

\begin{equation}\label{eqn:NRB_indicator_func}
I_{RB}(i,j)=
    \begin{cases}
      1, & \text{if } j \in k(i),~i\in [1, \mathcal{N}], \;~j \in [1, K],~i,j \in \mathbb{Z} \\
      0, & \text{otherwise}
    \end{cases}
\end{equation}

In (\ref{eqn:NRB_indicator_func}), $I_{RB}(i,j)=1$ when the $j$-th RB is used by the $i$-th task. Further, the utilization of the RB for $i$-th task is
\begin{equation}\label{eqn:uRB}
u_{RB}(i)= \dfrac{\sum_{j=1}^{K} I_{RB}(i,j)}{K} \cdot \mu(i)
\end{equation}
where $\mu(i)$ is the utilization of the RB because a portion of an RB may be used, defined as $\mu(i) = \dfrac{ T_{cm}(i) }{ T_p }$.

Because there is no overlapping of RB allocation among tasks, we can also see that
\begin{equation}\label{eqn:Tasks_IRB}
 \sum_{i=1}^{\mathcal{N}}\sum_{j=1}^{K} I_{RB}(i,j) = K
\end{equation}

In a case where each task is assigned with equal number of RBs, we can have $|k(i)| = c \cdot K/\mathcal{N}$, $c \in \mathbb{R^+}$ is a constant ratio, if $\mathcal{N}  \leq K$, $0 < c < K-\mathcal{N}$, as we want to ensure there is at least one RB for a task; while if $\mathcal{N} > K$, $0 < c \leq 1$. 

We can then generally formulate the optimization problem as follows, where the objective is to maximize the resource utilization for all given tasks.

\begin{equation*}
\begin{aligned}
& \underset{i}{\text{Maximize}}
& & \sum_{i=1}^{\mathcal{N}} u_{c}(i) \cdot  u_{RB}(i) \\
& \text{Subject to}
& & \sum u_c(i) \leq 1, \; i = 1, \ldots, \mathcal{N}. \\
& & & \sum u_{RB}(i) \leq 1, \; i = 1, \ldots, \mathcal{N}.\\
\end{aligned}
\end{equation*}

Let us look at the $u_c(i)$ here. First, we need to check the task execution at M to see how the FDDS runs in terms of compute resource utilization. In Fig. \ref{FigCommComp}, there are multiple tasks that utilize the compute resource on M, where the time period for the $i$-th task is split into three types: idle, communication, and computation, denoted by $T_{idle}(i)$, $T_{cm}(i)$, and $T_{cp}(i)$, respectively. We can see that for each time slot $T_p$, we have $T_p \geq T_{idle}(i) + T_{cm}(i) + T_{cp}(i)$. Due to the fact that the time taken on computation is negligible compared to the communication and M is usually equipped with a high-performance computing resource, for each time slot $T_p$, we can assume $T_{cp}(i) \ll T_{idle}(i) + T_{cm}(i)$ and $\sum T_{cp}(i) \leq T_p$. If we argue that during $T_p$, all $\mathcal{N}$ tasks will be executed, we have $\sum^{\mathcal{N}} T_{cp}(i) \leq 1$ . Besides, we can observe that, in an extreme case, for the $i$-th task, if $T_{idle}(i) + T_{cm}(i) \geq T_p$, then $T_{cp}(i) = 0$, meaning that a task will not get an opportunity to be executed in the given time slot.

Now we can prove that the constrained optimization problem formulated is NP-complete.

\begin{IEEEproof}
The formulated optimization problem is reducible to the weighted version of the set packing problem which is NP-complete. We can express the objective function as
$\sum_{i=1}^{\mathcal{N}}\mathcal{W}(i)\cdot |k(i)|$, where $\mathcal{W}(i) = \dfrac{1}{K} \left(\dfrac{T_p}{T_{cm}(i)} + \dfrac{T_p}{T_{cp}(i)} \right)^{-1}$, which is less than one following the time slot allocation in (\ref{eqn:Tp}). The objective function can be further expressed as 
\begin{equation*}
\sum_{i \in \mathbf{N} } \mathcal{W}(i) \cdot |k(i)|
\end{equation*}
Thus, the optimization problem can be described as finding a set packing that uses most sets in $\mathbf{K}$. This is equivalent to the classical maximum set packing problem, which is NP-complete.

\end{IEEEproof}

The complexity of the solution to the problem is captured by the set packing problem, which is related to the RB allocation policy represented by a function $\lambda$. However, it has been proven that limiting the element size of a set packing, for example, in our case $|k(i)|=2$, the problem is solvable in polynomial time \cite{papadimitriou1994computational}. However, with the typical cases with additional assumptions that we will discuss in the subsequent subsections, the problem can be converted to optimization versions with equality constraints which can be solved in polynomial time with a Lagrange multipliers method.

\subsection{Virtual Processor Versus FDDS Task Execution }
In an actual deployment of an MEC server, an MEC server can be a virtual machine (VM) with compute resources for the FDDS tasks, where each VM has one or more virtual processors (VPs). If we consider FDDS tasks shown in Fig. \ref{FigCommComp} are executed on a VP, where the execution of tasks happens in a sequential fashion, we can better discuss the extensibility of the FDDS tasks. For example, for Task 1 shown in Fig. \ref{FigCommComp}, minimizing the time after $T_{cp}$ may be utilized for the streaming of another set of data as $T_{cm}$ does not occupy the processor time at all. However, we should be aware that the time after $T_{cp}$ for Task 1 may not be sufficient as there are other tasks to be executed.  This may make impossible the utilization of the idle time until the next cycle with the $T_p$ timeframe. However, if there are multiple VPs on an MEC server, we can implement this idea by scheduling that to occur on another VP. 

Now let us discuss the solution to the optimization problem based on the following typical cases. 

\subsection{Case I: Ideal VP and RB Utilization}
In an ideal case, we let $u_c(i)$ and $\mu(i)$ approximate to 100\%, respectively, i.e., $u_c(i)  \to 1$, $\mu(i) \to 1$. Based on (\ref{eqn:ucc}) and (\ref{eqn:uRB}), knowing that $u_c = \sum u_c(i) < 1$ and $\mu(i) \leq 1$,  we can derive the upper bound of $U_{cc}$
\begin{equation}\label{eqn:ucc_case1}
U_{cc}^{\text{I}}=\sum_{i=1}^{\mathcal{N}} u_{c}(i) \cdot  c/\mathcal{N} \cdot \mu(i) \leq c
\end{equation}
where $U_{cc}^{I} = c$ if $u_c(i) = 1$, $\mu(i) = 1$. This indicates the upper bound of $U_{cc}$ is related to the RB allocation strategy for a task.

\subsection{Case II: Multiple FDDS Tasks with An Ideal Scheduling}
In this case, we will consider a more specific scenario compared to Case I, where we let RBs be allocated equally to all tasks so the RB for each task is $c \cdot K/\mathcal{N}$, $c \in \mathbb{R^+}$, and we let $T_{cp}(i)$ be $T_{cp}$ for all tasks, and let $T_{cm}(i)$ be $T_{cm}$, and all tasks have the same values of $\mu(i)$. Therefore $T_p = c_0 + T_{cm} + \mathcal{N}\cdot T_{cp}$, which implies the satisfaction of the condition $\sum u_c(i) \leq 1, \; i = 1, \ldots, \mathcal{N}$.

The definition of $T_p$ in this case can be illustrated in Fig. \ref{FigCommComp} where the time slot is adaptive to the required time for delivering the required data to M and the computation time for the data. The value $c_0$ is a constant which can be considered as one or more intervals of a protocol.

We can express $U_{cc}$ for Case II as
\begin{equation}\label{eqn:ucc_case2}
\begin{matrix}
U_{cc}^{\text{II}}= c  T_{cp} \cdot \dfrac{T_{cm}}{(T_{cm}+ T_{cp}) \cdot \left( c_0 + T_{cm} + \mathcal{N} T_{cp}\right) } 
\end{matrix}
\end{equation}
where $T_{cp} >0$ and $T_{cm}>0$. The maximum value of $U_{cc}^{\text{II}}$ can be obtained when
\begin{equation}
T_{cm} = \sqrt{T_{cp} \left( \mathcal{N} T_{cp} + c_{0}\right)}
\end{equation}

\subsection{Case III: Multiple FDDS Tasks with A Realistic Scheduling}

Now let us see a general case, where we let $T_p=\mathcal{N} c_0+ T_{cm}$ and keep other conditions the same as Case II. The definition of $T_p$ corresponds to a protocol design where at M the computation job for a task is not dependent on $T_{cp}(i)$ but mainly on $T_{cm}$. As illustrated in Fig. \ref{FigCommComp}, through a padding of the time slot represented by $\mathcal{N}c_0$, we can make $T_p$ a constant time slot with the length of $\mathcal{N} c_0+T_{cm}$. 

We can express $U_{cc}$ for Case III as
\begin{equation}\label{eqn:ucc_case3}
\begin{matrix}
U_{cc}^{\text{III}}= c T_{cp} \cdot \dfrac{T_{cm}}{(T_{cm}+ c_2) \cdot \left(  \mathcal{N}c_0 + T_{cm} \right) }  
\end{matrix}
\end{equation}
where $T_{cp} >0$ and $T_{cm}>0$. For each $(T_{cm}, T_{cp})$, the maximum value of $U_{cc}^{\text{III}}$ is obtained when: 
\begin{equation}\label{eqn:ucc_Toptimum}
T_{cm} = \sqrt{\mathcal{N} c_0 \cdot T_{cp}}
\end{equation}
where the first derivative of (\ref{eqn:ucc_case3}) needs to meet the condition that $\mathcal{N} c_0 \neq T_{cp}$. 

\begin{figure}[!t]
\centering
\includegraphics[width=3.2in]{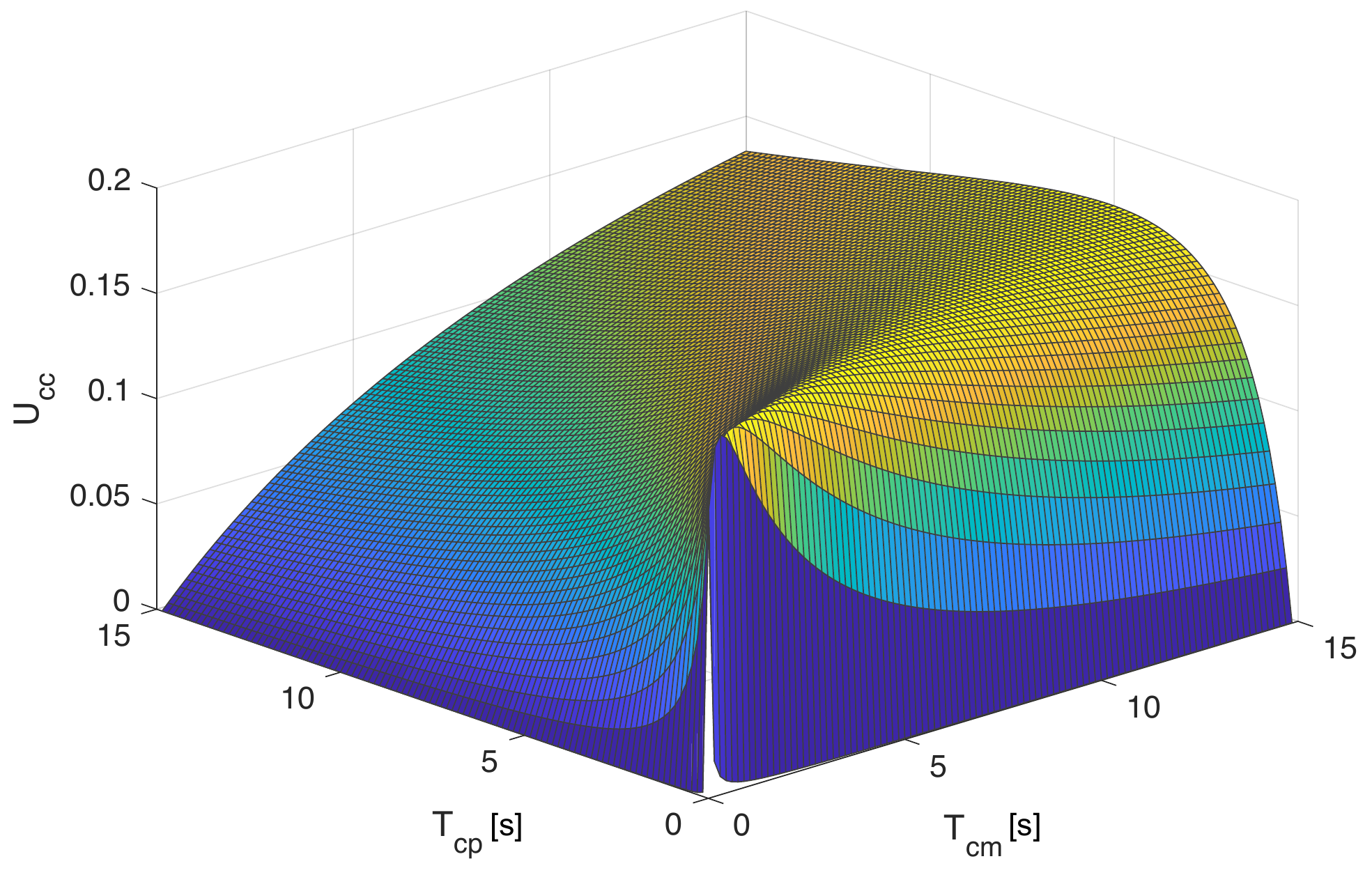}
\caption{Relationship of $T_{cm}$ and $T_{cp}$}
\label{FigTcommTcomp}
\end{figure}

From Fig. \ref{FigTcommTcomp}, we can see how $U_{cc}$ changes with various $T_{cm}(i)$ and $T_{cp}(i)$, where an optimal $U_{cc}$ is achievable with (\ref{eqn:ucc_Toptimum}). We can also see to obtain an optimal value of $U_{cc}$,  $T_{cp}(i)$ slowly increases in a non-linear fashion when $T_{cm}(i)$ increases.

\subsection{Discussion on Irregular {$T_{cm}$} for Tasks}
Let us see another general case that the amount of data $D$ is not the same but changeable with regard to a discrete random variable (r.v.) $\mathbf{a} = \{a_1, a_2,..., a_{\mathcal{N}}\}, \mathbf{a} \in \mathcal{A}$. For the $i$-th task, we let $D(i) = a_i \cdot D$. 

For multiple data flows where $T_{cm}(i)$ and $T_{cp}(i)$ are assumed positively proportional to the amount of data, if delivering $D$ takes $T_{cm}(0)$ and processing $D$ takes $T_{cp}(0)$, $T_{cm}(i)$ and $T_{cp}(i)$ will be $a_i \cdot T_{cm}(0)$ and $a_i \cdot T_{cp}(0)$, respectively. In this sense, we can see that $\mathbf{T_{cm}}$ and $\mathbf{T_{cp}}$ are r.v., and $\mathbf{T_{cm}} = T_{cm}(0) \cdot \mathbf{a}$,  $\mathbf{T_{cp}} = T_{cp}(0) \cdot \mathbf{a}$.

To get a closed-form expression, let us suppose $\mathbf{a} {\sim} N(\mu, \sigma^2)$. Then we have $\mathbf{T_{cp}} {\sim} N(T_{cp}(0) \mu$, $T_{cp}(0)^2 \sigma^2)$, and $\mathbf{T_{cm}} {\sim} N(T_{cm}(0) \mu, T_{cm}(0)^2 \sigma^2)$. Therefore, the expected values of $\mathbf{T_{cm}}$ and $\mathbf{T_{cp}}$ are $T_{cm}(0) \mu$ and $T_{cp}(0) \mu$, respectively. If we let $\hat{T}_{cm}(i) =  T_{cm}(0)\mu$ and let $\hat{T}_{cp}(i) =  T_{cp}(0)\mu$, we can still apply the analysis for Case II to Case III.

\subsection{Discussion on Practical Deployments}
Based on the framework we presented, we have revealed some principles of the CEC-based protocol designs. There are more cases can be defined. For example, how $U_{cc}$ will look like in a parallel computing scenario. 

There are practical considerations when the aforementioned results are applied into a real-world deployment. On the one hand, at the edge, it is possible to offload FDDS tasks to other VPs on M. We intend to discuss some basic cases before we can extend them to a more complex scenarios where multiple VPs and multiple FDDS tasks are present. This is being done in another paper where we try to find an optimum offloading strategy when using multiple VPs for FDDS-like tasks in a CEC loop. On the other hand, the limited RBs in a 5G network are basically shared by various tasks, which may cause scheduling conflicts for example at the radio link control (RLC) layer that ultimately affect the performance of FDDS tasks, in this case, we can consider the solutions of using federated radio resources, such as the Cloud RAN \cite{Checko2015}. A further discussion on this is out of the scope of the current paper.

\section{CEC-based Communication Protocol}

To apply the aforementioned results in Section IV into an FDD scenario, we propose the ReFlexUp protocol described in Alg. \ref{ReFlexUp}, where the procedures occur in key entities for each iteration. In Alg. \ref{ReFlexUp}, ReFlexUp is implemented on four entities: eNB node $\text{C}$, MEC server node $\text{M}$, relay node $s$, and sensor node $v$. C and M are assumed to be directly connected, and M can coordinate the use of the radio resources on C. The first step of determining the criteria is to be done on M, where M may have the prior knowledge of the RBs to be provisioned. M will then determine the CEC efficiency based on the Case III in Section IV. 

In Alg. \ref{ReFlexUp}, with each iteration, data preparation is required for a task through the URLLC, where each $s$ node will report the number of data flows to M in order to let M and C calculate the key parameters such as $c_0$, $c$, $|k(i)|$, and $K$, which will lead to the calculated results of $T_{cm}(i)$ and $T_{p}$. Then, each $s$ node will be informed of the parameter $T_{cm}(i)$ based on, for example, (\ref{eqn:ucc_case3}). Then, each $v$ node associated with an $s$ node needs to send the sensing data packets to the $s$ node first and then the data will be relayed to C. At this point, each $s$ node may initiate the possible packet re-transmission procedure if the required data received at C is less than that as specified by the CEC criteria $\varepsilon$. C will cache the previous packet and transmit the complete packet to C. ReFlexUp is configured to have a threshold $\varepsilon$ for considering the communication failure probability $p_{cf}$, where if the FDD process requires all data from local network to M, then $\varepsilon = 100\%$. For some industrial automation applications, the threshold $\varepsilon$ might be less than 100\% if some packet loss can be tolerated by a computation task at M.

At the beginning of Alg. 1, we can assume the RB resource is allocated a priori as for FDD tasks with a dedicated eNB, the RB allocation scheme may not need to change dynamically. However, it can be determined, for example, through a dynamic process made prior to a session of data stream transfers for each iteration, which can be integrated because ReFlexUp does not impose any limit on the RB allocation schemes. The other resources and configurations on nodes are assumed to be known during an initial planning phase which is a common practice in industrial application deployments. 

ReFlexUp requires the use of the optimal communication with the consideration of the computation time of the FDDS at M. Relay nodes are the ones directly interacting with C and they need to meet such a requirement with two possible strategies: the $s$ nodes can adapt to adjust the rate $R$ for decreasing the $T_{cm}$, or the $s$ nodes can reduce the amount of data to be transmitted per relay node by increasing the ratio of number of $s$ nodes to the number of $v$ nodes.

\begin{algorithm}
    \caption{ReFlexUp}
    \label{ReFlexUp}
    \begin{algorithmic}[1] 
        \Procedure{ Prepare data }{$s$, $v$}
        	\State $s \gets S, v \gets V$, $S,V \subset G$
            \State for each $s_i$, $s_i \in S$
            	\Procedure{Calculate the parameters}{$N$, $T_p$ and $T_{cm}(i)$ bounds}
				\State $s_i$ reports the number of data flows to M
				\State Initialize parameters (e.g., $c_0$, $c$, $|k(i)|$, $K$, etc.)
				\State Calculate $T_{cm}(i)$ and $T_{p}$
				\State Inform each $s_i$ of $T_{cm}(i)$
				\State Each $s_i$ makes a schedule for communication with its member nodes in $V$
			\EndProcedure
			\If{There exists a set of nodes $\{v_{i,j}\}  \subset V$ associated to $s_i$}
			\State Each node $v_{i,j}$ sends data to $s_i$ with the allocated RB
			\Else
				\State $s_i$ sends data packets $D_i$ to C with the allocated RB
				\Procedure{ Determine the CEC criteria}{$\varepsilon$}
				\If{the data received at C equals $\varepsilon$}
				\State An FDDS task at M processes the data
				\Else
				\State An FDDS service caches the current packets and notify the $s_i$ of the list of missing packets $\hat{D_i}$
				\Procedure{Packet Re-transmission Scheme}{$\hat{D_i}$}
				\State $s_i$ transfers the missing packet with the previously cached packet to C
				\EndProcedure
				\EndIf
				\EndProcedure
			\EndIf
            
        \EndProcedure
    \end{algorithmic}
\end{algorithm}

ReFlexUp is different from the existing communication protocols at least with the following features. (1) It is specifically designed for the FDD scenarios with the proposed CEC loops in a 5G and MEC environment. (2) It utilizes the results of the proposed system model with the optimized CEC efficiency considering the essential architectural entities. Aligned with this feature, ReFlexUp is adaptive to the radio environments in terms of the fading and signal-to-noise ratio (SNR) where the member nodes are informed of the optimal $T_{cm}(i)$ in (\ref{eqn:ucc_Toptimum}). The determination of the $T_{cm}(i)$ occurs during each $T_p$ time period. (3) It is network transport agnostic which can be implemented at a higher layer on top of different transports. (4) It incorporates a relay transmission to combat against the complex radio environment in the industrial settings. Since ReFlexUp does not specify the local communication between nodes in $V$ associated to a relay node in $S$, this can allow ReFlexUp to work on top of various underlying network protocols used if the latency and reliability performance are not affected. For example, the existing ARQ, Selective Repeat ARQ, and HARQ protocols do not have the listed features, and the Occupy CoW protocol does not have the first three features. 

Let us look at the computational complexity of ReFlexUp. In the first procedure, each $s$ node first reports its number of data flows to M in order to let M determine the parameters for each data flow based on the results in Section IV. The computational complexity of this process at M is generally based on the complexity of the solutions to the optimization problem formulated in Section IV. However, from (\ref{eqn:ucc_case3}), we know that it can be a constant value $y_0$ as the calculation to solve (\ref{eqn:ucc_case3}) occurs once and ReFlexUp only needs to use the results of the derivative test shown in (\ref{eqn:ucc_Toptimum}). Next, during the data transmission procedure, if we suppose one transmission of data packets from $v$ to $s$ and from $s$ to M take 1 unit of time, respectively, the transmission of data for each $s_i$ takes $|V_i|$ + 1, and for all nodes in $S$, the time complexity will be $n_s \cdot (|V_i| + 1)$. Suppose each $s_i$ is associated to $\frac{n_v}{n_s}$ sensor nodes, that becomes $n_s \cdot (\frac{n_v}{n_s} + 1)$. The third procedure is greedy based at M, where, in the worst case, it will take $ \varepsilon \cdot D_i / m$, as each $s_i$ can transmit in parallel sessions due to the fact that each $s$ node has individual communication resources allocated. Suppose each $s_i$ has equal length of data, i.e., $D_i=D$. It becomes $ \varepsilon \cdot D/ m$. To summarize, the computational complexity in these procedures for $\mathcal{N}$ tasks is $\mathcal{N} \cdot (y_0 + n_s \cdot (\frac{n_v}{n_s} + 1)$ + $ \varepsilon \cdot D / m )$. If we use the big O notation, the complexity of ReFlexUp is O$(\mathcal{N} \cdot (n_v + n_s))$.

\section{Performance Analysis}
We will compare the proposed ReFlexUp protocol with other protocols, i.e., the Selective Repeat ARQ protocol, generic HARQ protocol, and the relay-based Occupy CoW. The Selective Repeat ARQ protocol can be practically used in various communication systems including the IEEE 802.11n wireless networks. The generic HARQ is a typical protocol broadly used in LTE and cellular IoT systems. We will discuss $T_{cm}$ and the probability of transmission failure (or outage) $P_{fail}$ of the protocols.

There are a few assumptions we need to make in order to compare the protocols. First, as $T_{cm}(i)$ denotes the total amount of time taken to get the required data $D(i)$ for the $i$-th task, following the assumptions made in Case III of Section IV, we assume each FDDS task requires the same amount of data, i.e., $D(1)=...=D(i)=...=D, 1 \leq i \leq \mathcal{N}$. Thus, each of $\mathcal{N}$ tasks needs $D$ bits of data to perform the computation job within a timeframe $T_p$. $D$ is split into small packets (where each packet has $m$ bits) sent by  sensor nodes. If there is a decoding/transmission error occurred on a packet, it needs to be individually retransmitted again. Further, the Rayleigh fading and additive white Gaussian noise channel are used for the analysis.

We use the definition of link failure in \cite{Swamy2017}, which is the outage probability in the Rayleigh fading channel:

\begin{equation}\label{eqn:generic_pl}
P_{l} = P(C<R) = 1 - exp \left(-\frac{2^{\frac{R}{W}} - 1}{\text{SNR}} \right)
\end{equation}
where $W$ is the bandwidth and $R$ is the rate. 

\subsection{Selective Repeat ARQ Protocol}
Following the classical Selective Repeat ARQ, we do not use relay nodes so all nodes in $V$ can directly transmit data packets to C. If the probability of the delivery failure of a packet is considered the same as the link failure $P_l$ in (\ref{eqn:generic_pl}), and suppose the average retransmission round is 1. Knowing that the packet delivery failure due to timeouts or packet errors will result in a retransmission, and  assuming there are no decoding errors, we can have the average communication latency $T_{cm}(i)$ as follows:

\begin{equation}
T_{cm} = \dfrac{ N_v \cdot m \left( 3 - 2P_{l} \right) } {R}
\end{equation}

The failure of the ARQ-based packet transmissions is:
\begin{equation}\label{eqn:pfail_sereparq}
P_{fail}=p_{a} + (1-p_a) \cdot p_b
\end{equation}
where here $p_a $  is the probability of the event when a transmission timeout occurs, and $p_b$ is the probability of the event when a packet error occurs.

\subsection{HARQ Protocol}
In HARQ, decoding errors need to be considered compared to the Selective Repeat ARQ, so $P_f$ is replaced with the outage probability that a packet fails to be decoded after $Q$ HARQ rounds \cite{Wu2011}:
\begin{equation}
P_{fail} = \mathbb{P} \left(  \sum_{i=1}^{Q}  \left( \dfrac{1}{L} \sum_{j=1}^{L} \text{log}_2(1 + \text{SNR} |h_{i,j}|^2) \right)  \leq R   \right)
\end{equation}
where $R$ is the data rate and $L$ is the diversity order. The communication latency is dependent on the average rounds given in (3) of \cite{Wu2011}. If $Q=1$, HARQ is reverted to the stop-and-wait ARQ. 

From \cite{Wu2011}, the closed-form expression of the approximated expected number of packet retransmission rounds $\hat{d}$ is given, based on which we can obtain the average $T_{cm}$ can be expressed as $ \frac{ \hat{d} \cdot N_G  m}{R}$.

\subsection{Occupy CoW based Protocol}
Occupy CoW \cite{Swamy2017} is based on the 2-phase cooperative relaying scheme, where if we denote the link failure of each phases by $p_1$ and $p_2$, respectively, and denote the link failure occurs in the 2nd phase given it fails in the 1st phase by $p_{12}$, the UL system failure in a Rayleigh fading channel with a fixed scheduling is given as follows:
\begin{equation}
P_{fail}=\sum_{a=1}^{n-1} \left\{ \binom{n}{a}  ( 1 - (1-p_1)^a + (1-p_1)) ( 1 - (1-p_{12}^{n-a} )   \right\}  
\end{equation}
where $a$ is the number of nodes successfully transmitted packets to the relay nodes. The overall time period for transmitting packets $T_{cm}$ consists of two parts, $T_1$ and $T_2$, where $T_1$ is the time taken for transmitting packets in the 1st phase and $T_2$ is the time taken for transmitting packets in the 2nd phase. In this case, we have
$p_1 = 1 - exp \left(-\frac{2^{\frac{n_v(m+1)}{T_1}} - 1}{\text{SNR}} \right)$, and $p_2 = 1 - exp \left(-\frac{2^{\frac{n_v(m+1)}{T_2}} - 1}{\text{SNR}} \right)$, and $p_{12} = min \left(\frac{p_1}{p_2}, 1 \right)$

\subsection{ReFlexUp Protocol}
We let the number of sensor nodes in $V$ associated to the $s_i$ node be $n_{i,v}$, and the rate of $s_i$ be $R=\frac{m(n_{i,v} + 1)}{ T_{v \rightarrow s} }$. The probability of a link failure based on (\ref{eqn:generic_pl}) is $P_l=1-exp(\frac{2^{\frac{m(n_{i,v} + 1)}{ T_{v \rightarrow s} }} - 1}{T_{v \rightarrow s}})$. The overall transmission failure $P_{fail}$ is dependent on the two-phase transmissions where each phase is similar to that of the Selective Repeat ARQ protocol (\ref{eqn:pfail_sereparq}) if ReFlexUp uses the ARQ-based protocol for data transmission in each phase.

\section{Performance Evaluation}

Here we make a general case for the FDDS tasks based on Case III in Section IV and the discussion in Section VI. We assume there are multiple FDDS tasks and each has the same requirements for $U_{cc}$ and $\varepsilon$. The efficiency is measured by the proposed $U_{cc}$ and the latency is measured by $T_{cm}$, where the reliability is measured by $P_{fail}$.

\subsection{Numerical Results}
Based on the previous analysis, we can explore the key performance metrics of ReFlexUp and compare them with the typical protocols. The numerical results are obtained based on the assumptions and numerical analysis in Section VI about latency, $U_{cc}$, and $P_{fail}$ with GNU Octave. Here we assume the protocols are at the Open Systems Interconnection (OSI) layer 1 and layer 2 which do not use higher layer features, so the numerical results can mostly reflect the actual protocol performance in essential scenarios. 

\begin{table}[H]
\caption{Parameters Used for Evaluation}
\label{Tbl:Parameter}
\centering
\begin{tabular}{c | c}
\hline
\textbf{Parameter} & \textbf{Value}\\ 
\hline
Bandwidth  & 20 MHz \\
SNR & [10 dB, 60 dB] \\
Packet size $m$ & 22 bytes\\
Rate & 200 kbps\\
$L$ & 2\\
$c_0$ & 1.5\\
$p_a$ & 0.0001\\
$T_{cp}$ & (0, 0.5] \\
$\varepsilon$ & 100\% \\
$n_s / n_v$  & 0.2 \\ 
$\mathcal{N}$ & 100 \\
\end{tabular}
\end{table}

The network is considered to be deployed in a factory cell where the nodes in $V$ are distributed and co-located with each other, and these nodes have direct connection to the nodes in $S$. The parameters used are shown in Table \ref{Tbl:Parameter}. Further, for ReFlexUp, we let the data rate of the two phases be the same, and we let $\varepsilon$ be 100\% to represent a strict requirement for an FDDS application. For Selective Repeat ARQ, we let the timeout probability $p_a$ be 0.0001, which is a very small value. For HARQ, we let the maximum of rounds with decoding failure be 7 and let the diversity order $L$ be 2, although we found that the value of $L$ will not affect the overall results. 

We will discuss the results for a generic FDDS scenario so we use the definition of $U_{cc}$ of the Case III in Section IV. We will evaluate the performance of theses protocols in terms of the CEC efficiency and latency.

\begin{figure}[!t]
\centering
\includegraphics[width=3.01in]{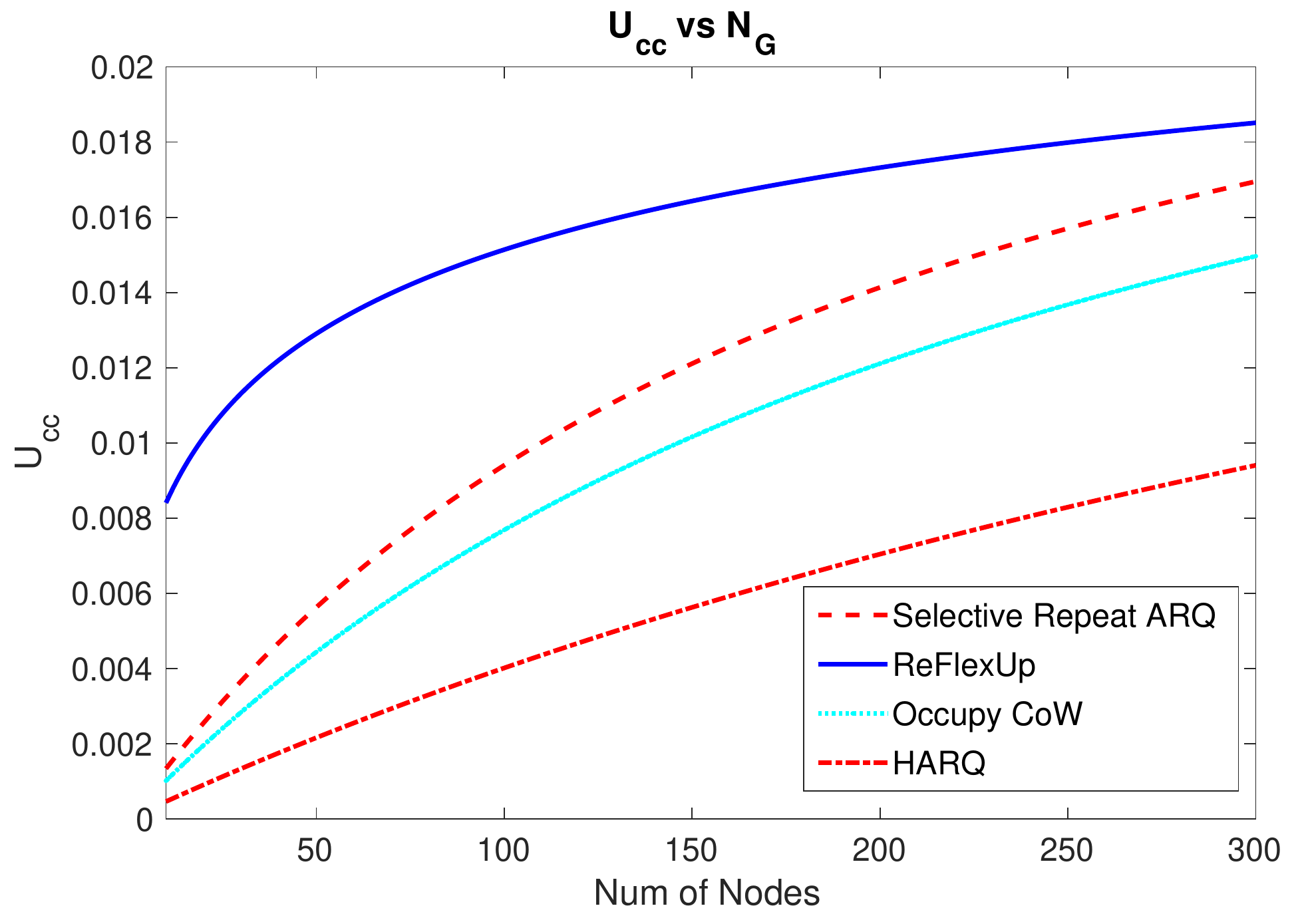}
\caption{$U_{cc}$ results of Selective Repeat ARQ, HARQ, ReFlexUp, and Occupy CoW}
\label{FigUccResult1}
\end{figure}

\begin{figure}[!t]
\centering
\includegraphics[width=3.01in]{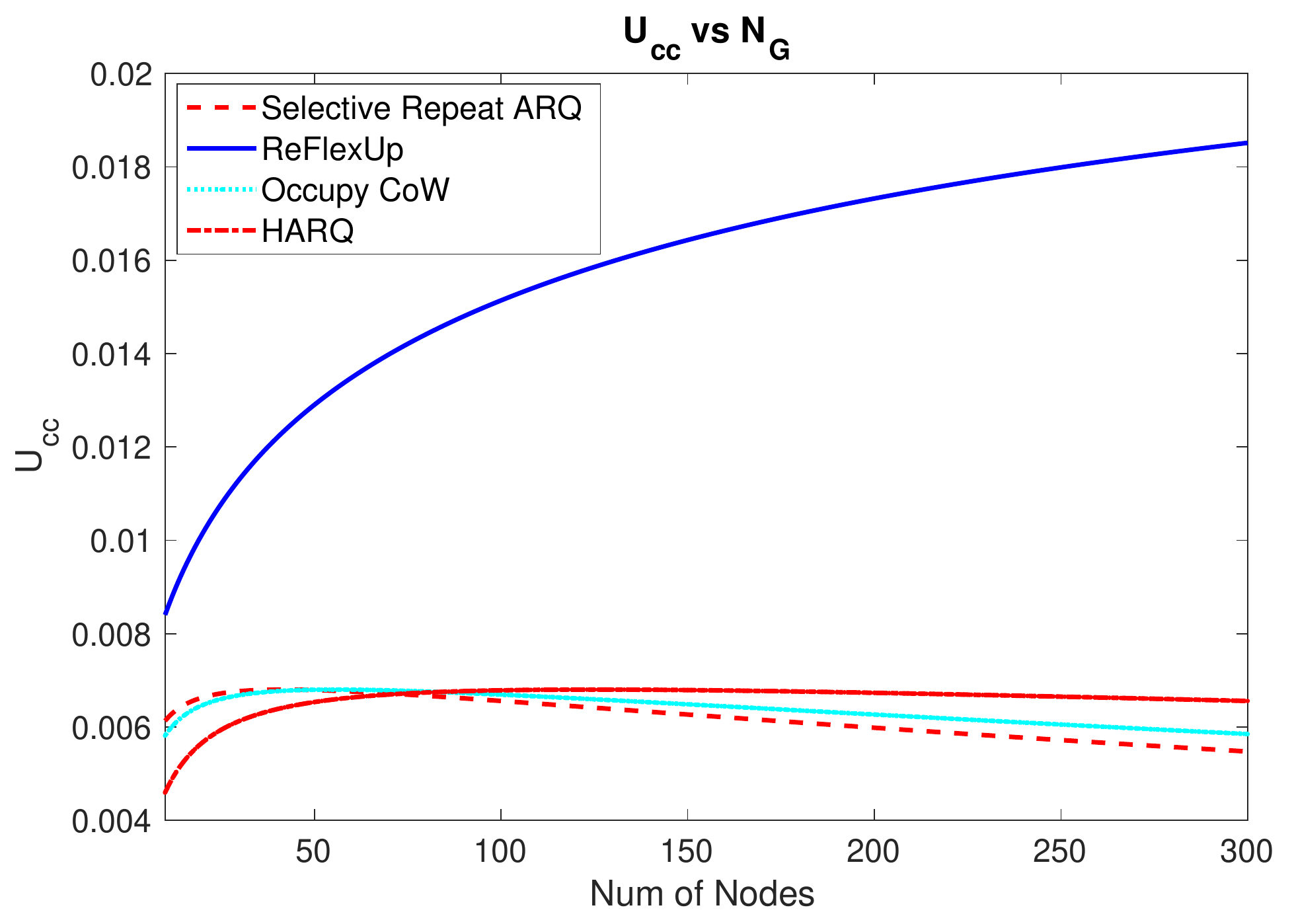}
\caption{$U_{cc}$ results of Selective Repeat ARQ, HARQ, ReFlexUp, and Occupy CoW}
\label{FigUccResult2}
\end{figure}

The results of the efficiency $U_{cc}$ are shown in Fig. \ref{FigUccResult1} when the maximum value of $T_{cp}$ is 0.5 s, where we can see that with the increasing number of nodes, ReFlexUp outperforms other protocols. This is because ReFlexUp uses the optimal $T_{cm}$ for each cycle of the FDDS task based on (\ref{eqn:ucc_Toptimum}). When the maximum value of $T_{cp} = 0.005$ s, the result is shown Fig. \ref{FigUccResult2}, where we can see ReFlexUp keeps the best efficiency while the performance of other protocols is degraded as the $T_{cp}$ setting affects the reliability of delivering sufficient data through a data transfer session. 

\begin{figure}[!t]
\centering
\includegraphics[width=3.01in]{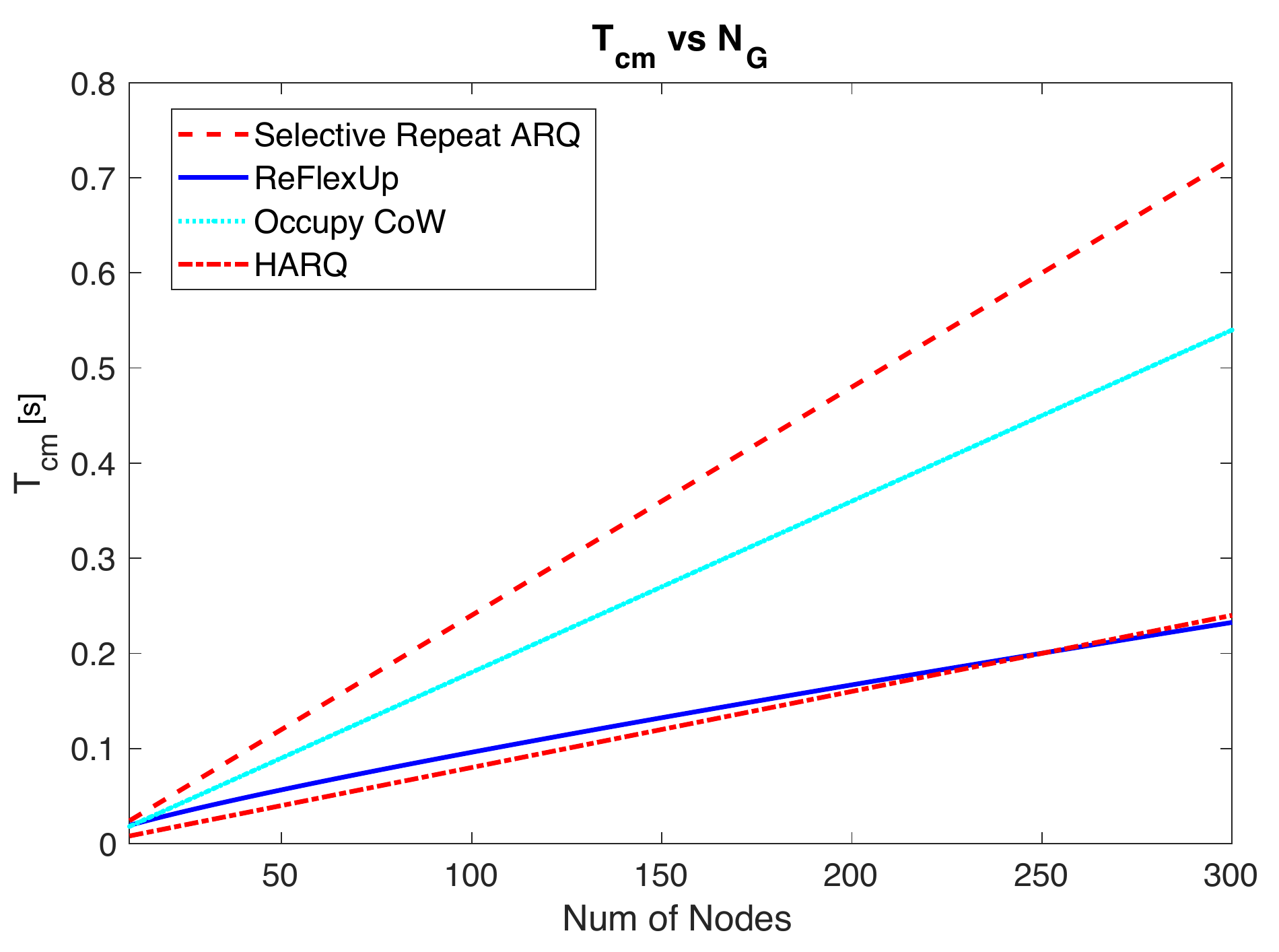}
\caption{$T_{cm}$ results of Selective Repeat ARQ, HARQ, ReFlexUp, and Occupy CoW}
\label{FigTcommResult}
\end{figure}

Fig. \ref{FigTcommResult} shows the performance of $T_{cm}$ versus the increasing network size, where the $T_{cm}$ value of ReFlexUp is similar to that of HARQ. This is because ReFlexUp uses a two-phase communication which causes a slight cost compared to that of HARQ. However, from $N_G = 251$, ReFlexUp outperforms HARQ, which shows the good performance of ReFlexUp when the network size scales up. In addition, from Fig. \ref{FigTcommResult}, we can see that HARQ takes much less time than that of Selective Repeat ARQ for all network sizes. This is because HARQ uses an error-correcting coding which reduces the required number of rounds of packet retransmissions. 

\begin{figure}[!t]
\centering
\includegraphics[width=3.01in]{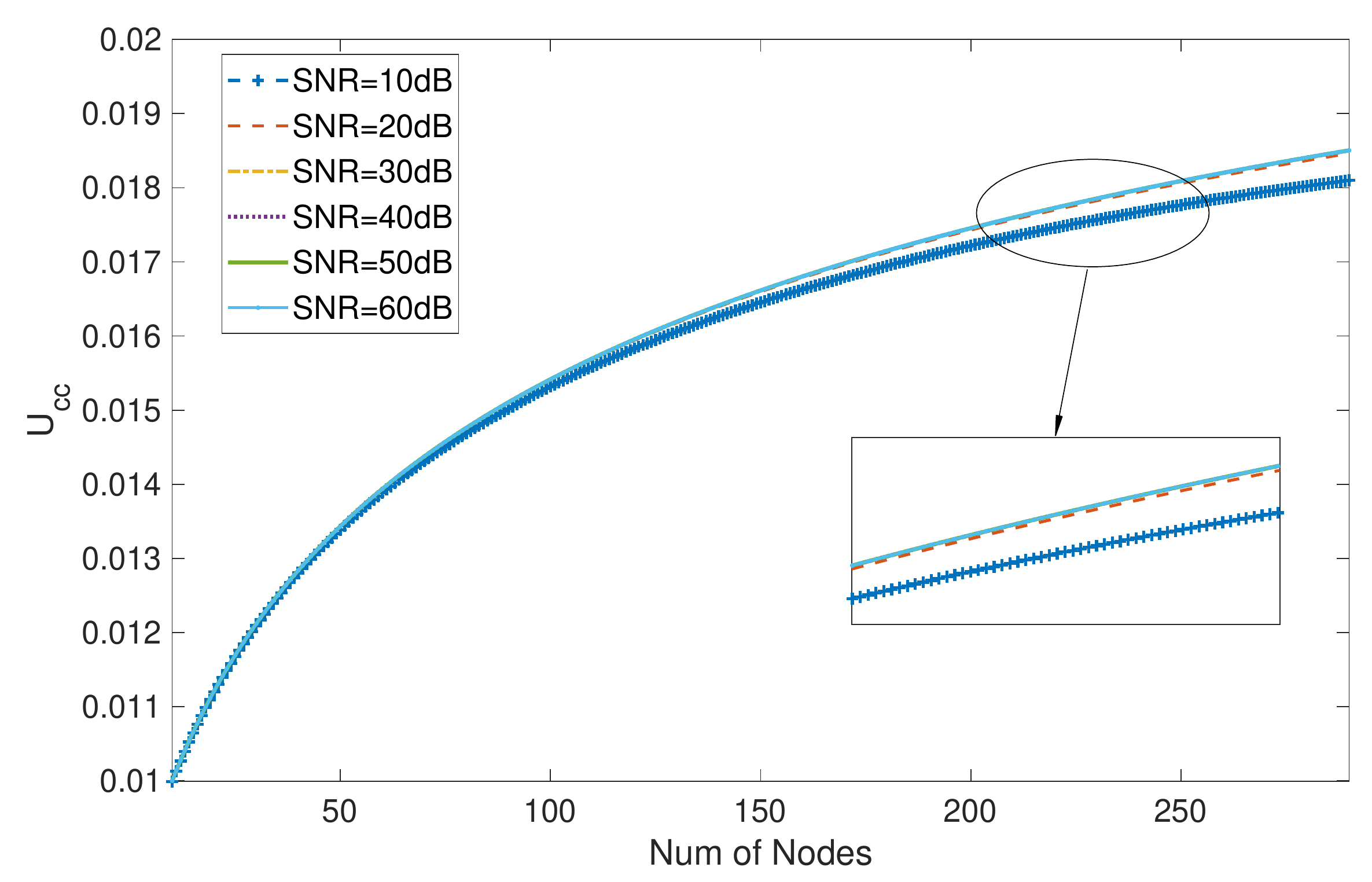}
\caption{$U_{cc}$ under various SNR conditions}
\label{FigUccResult3}
\end{figure}

The efficiency of ReFlexUp under different SNR conditions is shown in Fig. \ref{FigUccResult3}, where the $U_{cc}$ values are not  significantly affected by the SNR conditions event the SNR as poor as 10 dB. When the SNR value increases, $U_{cc}$ of ReFlexUp increases accordingly, although we can see that when the SNR is greater than 30 dB, $U_{cc}$ will only increase very slightly. This is because the SNR condition is already very good and it is sufficient to meet the communication requirements.

\begin{figure}[!t]
\centering
\includegraphics[width=3.01in]{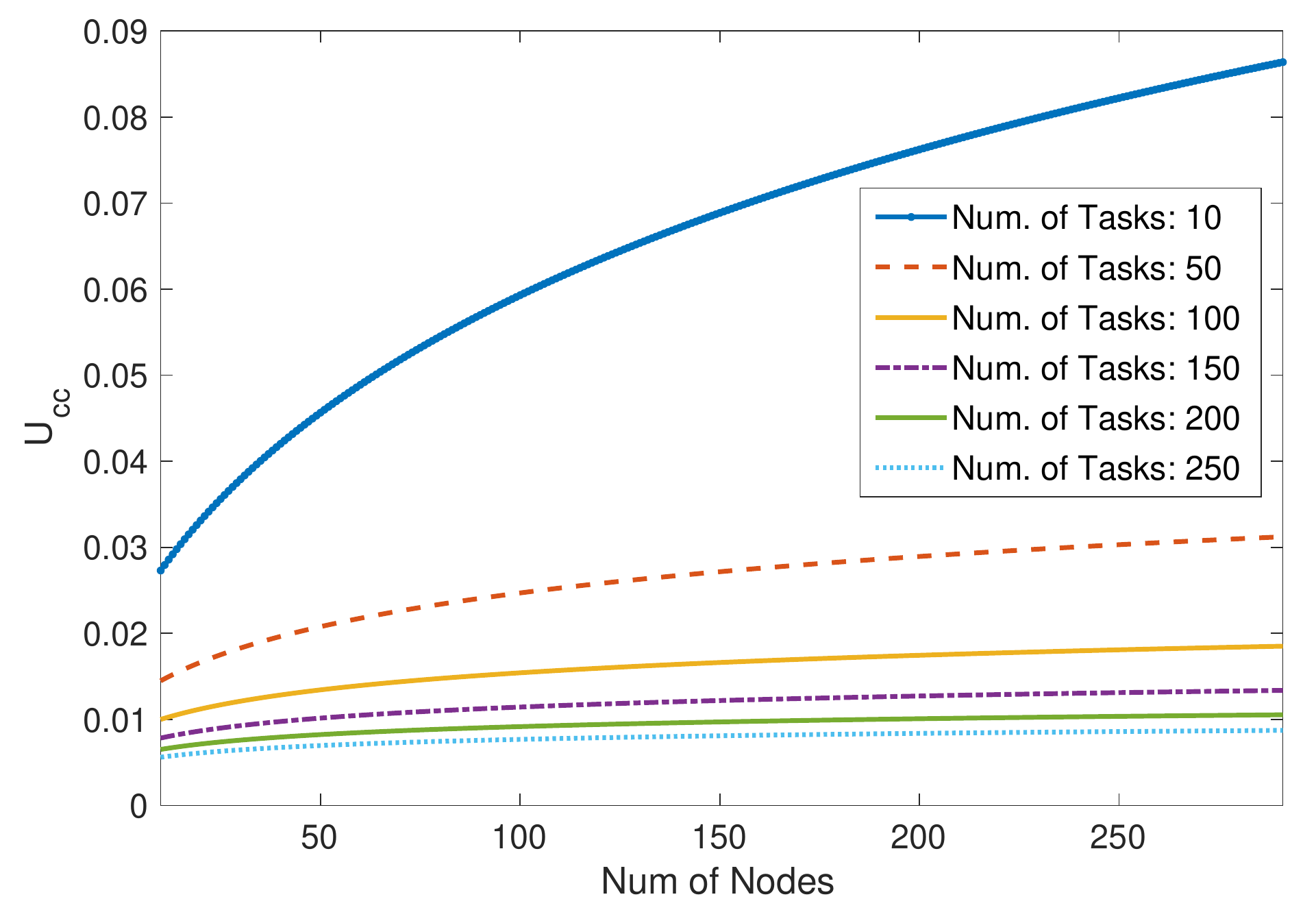}
\caption{$U_{cc}$ versus different number of FDDS tasks}
\label{FigUccResult4}
\end{figure}

The efficiency of ReFlexUp versus different number of FDDS tasks on a VP is shown in Fig. \ref{FigUccResult4}, where we can see that with the $U_{cc}$ is lowered when the number of tasks increases. In this sense, $U_{cc}$ has the best performance when the number of tasks is 10. This is because additional tasks require extra time on a single VP that affects the FDDS execution time and $T_{cm}$, which finally results in less efficient FDDS tasks. This also indicates that to achieve the preferable $U_{cc}$ performance, we need to do some computation offloading in order to keep the number of tasks running on a VP at M at a reasonable level. 

\begin{figure}[!t]
\centering
\includegraphics[width=3.01in]{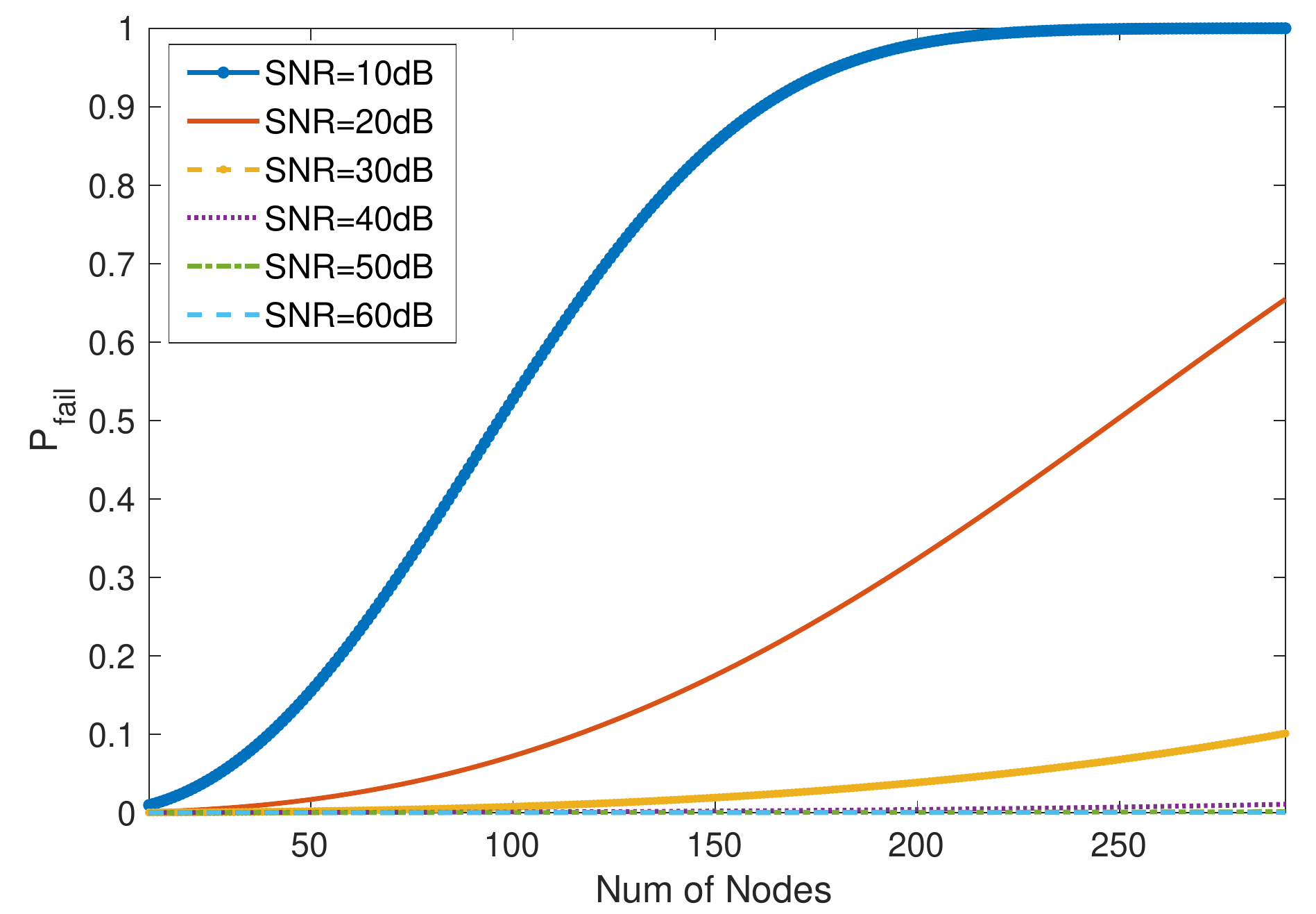}
\caption{$P_{fail}$ under various SNR conditions}
\label{FigPFailResult}
\end{figure}

At last, let us evaluate the reliability of the ReFlexUp in terms of $P_{fail}$. In Fig. \ref{FigPFailResult}, we can see that ReFlexUp can achieve very low $P_{fail}$ when the SNR is greater than 40 dB which usually referred as a good SNR condition. For example, in Fig. \ref{FigPFailResult}, if $N_G=250$, when SNR=40 dB, $P_{fail}=0.00708$, and when SNR=60 dB, $P_{fail}=0.00017$. These results indicate that in a good SNR condition, ReFlexUp can achieve the requirement of URLLC and at the same time achieve the efficiency of FDDS tasks with an optimal value of $U_{cc}$.

\subsection{Simulation Results}

In order to evaluate a generic FDDS application in a real deployment, we adopt the popular ns-3 network simulations to validate the proposed scheme in a real-world scenario, where a 5G communication stack is used on the nodes. Based on the Fig. \ref{Fig5GMecDeploy}, we consider the network model with mobile nodes are deployed in the 280 $\times$ 280 $\text{m}^2$ plane, where two relay nodes (i.e., UEs) are randomly distributed and each UE chooses the best eNB to connect. The MEC server is considered a host connected through the S1 interface to the EPC with a 100Gbps-bandwidth low-latency P2P link to an eNB. This P2P link is aligned with the architecture discussed in Section II and shown in Fig. \ref{Fig5GMecDeploy}, where the connection between MEC and eNB is assumed to have high bandwidth and extremely low latency. An FDD application is implemented as a UDP application with a server endpoint running on the MEC host and multiple client endpoints running on the UE nodes, where each UE is assumed to connect to multiple sensors and transmit data at a 50 $\mu$s interval. The extension of the setup can work with the 5G SA (standalone) deployment standardized in June 2018 \cite{Intel2018} and 5G non-standalone deployment with the co-existence of LTE and 5G RANs. In reality, the MEC server itself can be implemented with the mature OpenStack platform or Docker container platform provisioned through a 5G slicing process. The MEC server itself can run standalone and, can be alternatively be integrated to the eNB in the 5G RAN or to S/P-GW in the EPC. 

\begin{figure}[!t]
\centering
\includegraphics[width=3.10in]{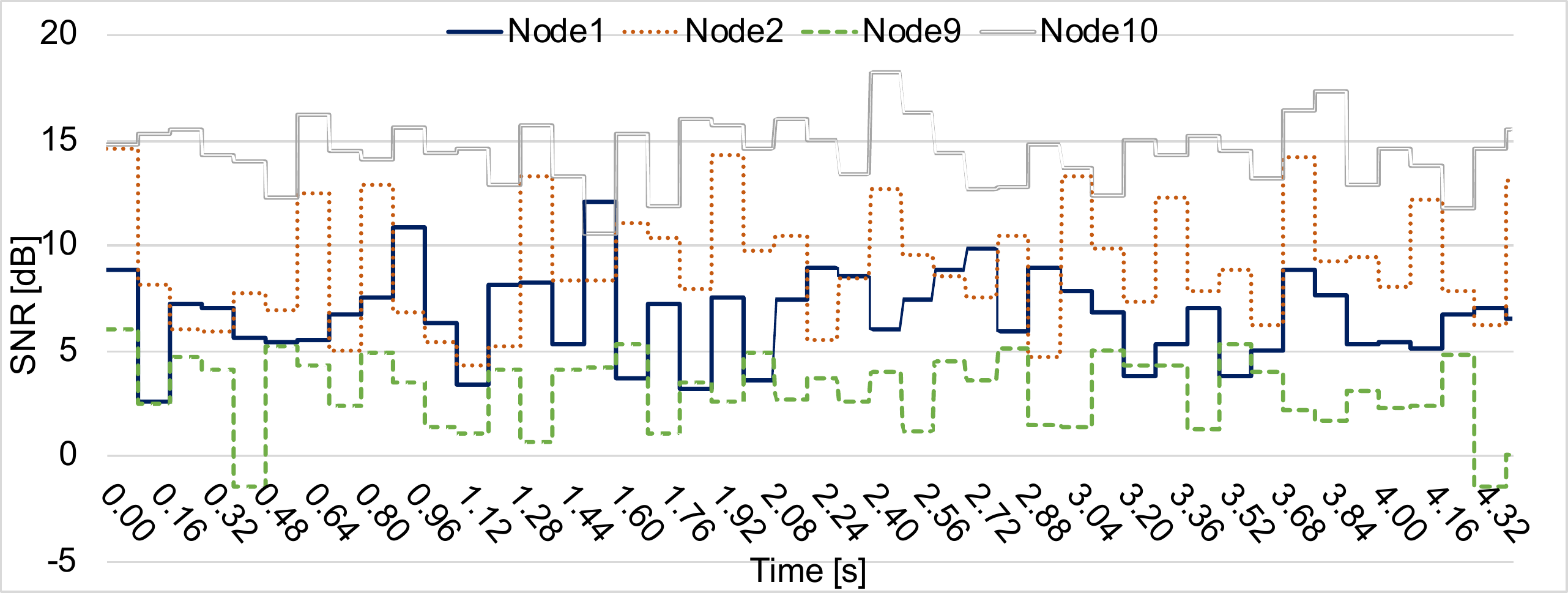}
\caption{Example SNR condition over time on uplink from mmWave UE nodes}
\label{FigSimuSNR}
\end{figure}

The ns-3 mmWave module is used in the experimentation, which utilizes the standardized beamforming and modulation coding scheme (MCS) level to determine the transport block size and subframe slots based on the NYU channel model \cite{MacCartney2014} with central frequency at 28 GHz. We evaluate the UL performance between ReFlexUp, Occupy CoW, and LTE, where the first two schemes are based on the 5G mmWave UL with 24 OFDM symbols per subframe. We assume the sensor nodes are pre-connected to a UE and the communication process is considered a retransmission process based on the FDDS task requirements. The default setting of the mmWave module and the parameters shown in Table \ref{Tbl:Parameter} are used, while a few key additional parameters are shown in Table \ref{Tb2:SimuParameter}. The 3GPP non-line-of-sight channel condition is used to best simulate a scenario shown in Fig. \ref{Fig1}. In the simulation, the SNR condition on a UL is temporally variable in relation to the location and radio environment as shown in Fig. \ref{FigSimuSNR} where we can see the distance between a UE and an eNB matters to the SNR condition and it also justifies the reason we use two mmWave eNBs in this simulation so that each node can connect to the eNB to have a reasonable channel condition.

\begin{table}[hbt]
\caption{Simulation Parameters}
\label{Tb2:SimuParameter}
\centering
\begin{tabular}{c | c}
\hline
\textbf{Parameter} & \textbf{Value}\\ 
\hline
mmWave eNB location & (50, 70, 3), (150, 70, 3) \\
LTE eNB location & (0, 0, 3) \\
Symbols per subframe  & 24 \\
Symbol period & 4.16 $\mu${s} \\
Subframe period & 100 $\mu${s} \\
RLC Tx Buffer Size & 20 MB \\
S1 latency & 1 ms \\
\end{tabular}
\end{table}

\begin{figure}[!t]
\centering
\includegraphics[width=3.01in]{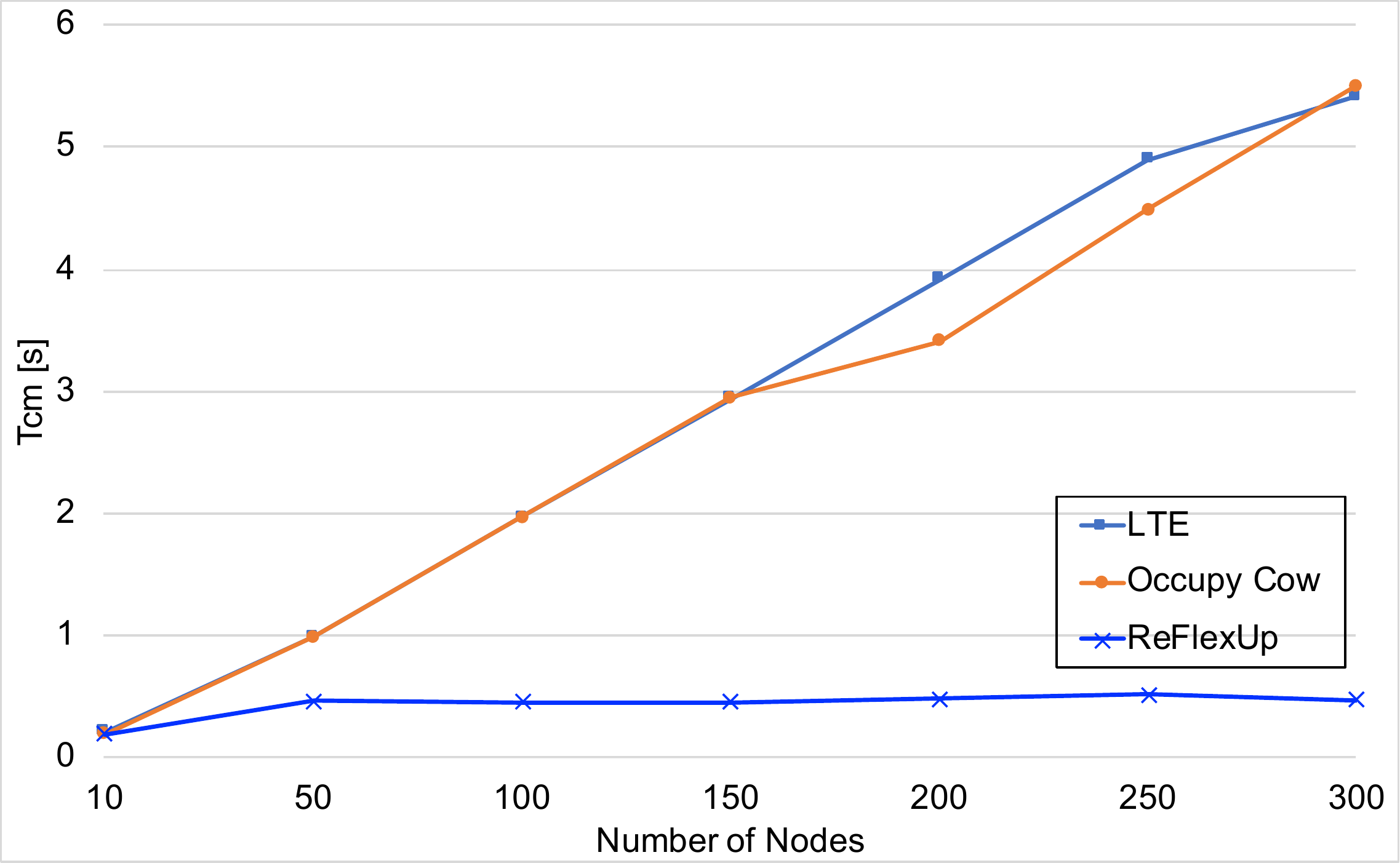}
\caption{$T_{cm}$ results of LTE, Occupy CoW, and ReFlexUp}
\label{FigSimuTcm}
\end{figure}

To evaluate the uplink performance in all protocols, we include the standard LTE as it has the typical HARQ mechanism, and for Occupy CoW, a UE node relays the data packet from each sensor node. In Fig. \ref{FigSimuTcm}, we use the incremental data amount to be sent from each $v$ node, we we can see ReFlexUp keeps the lowest $T_{cm}$ while Occupy CoW has a slightly lower $T_{cm}$ compared to that of LTE. This is because ReFlexUp can adjust its transmission strategy to meet the $T_{cm}$ requirement for an efficient edge task execution.

\begin{figure}[!t]
\centering
\includegraphics[width=3.01in]{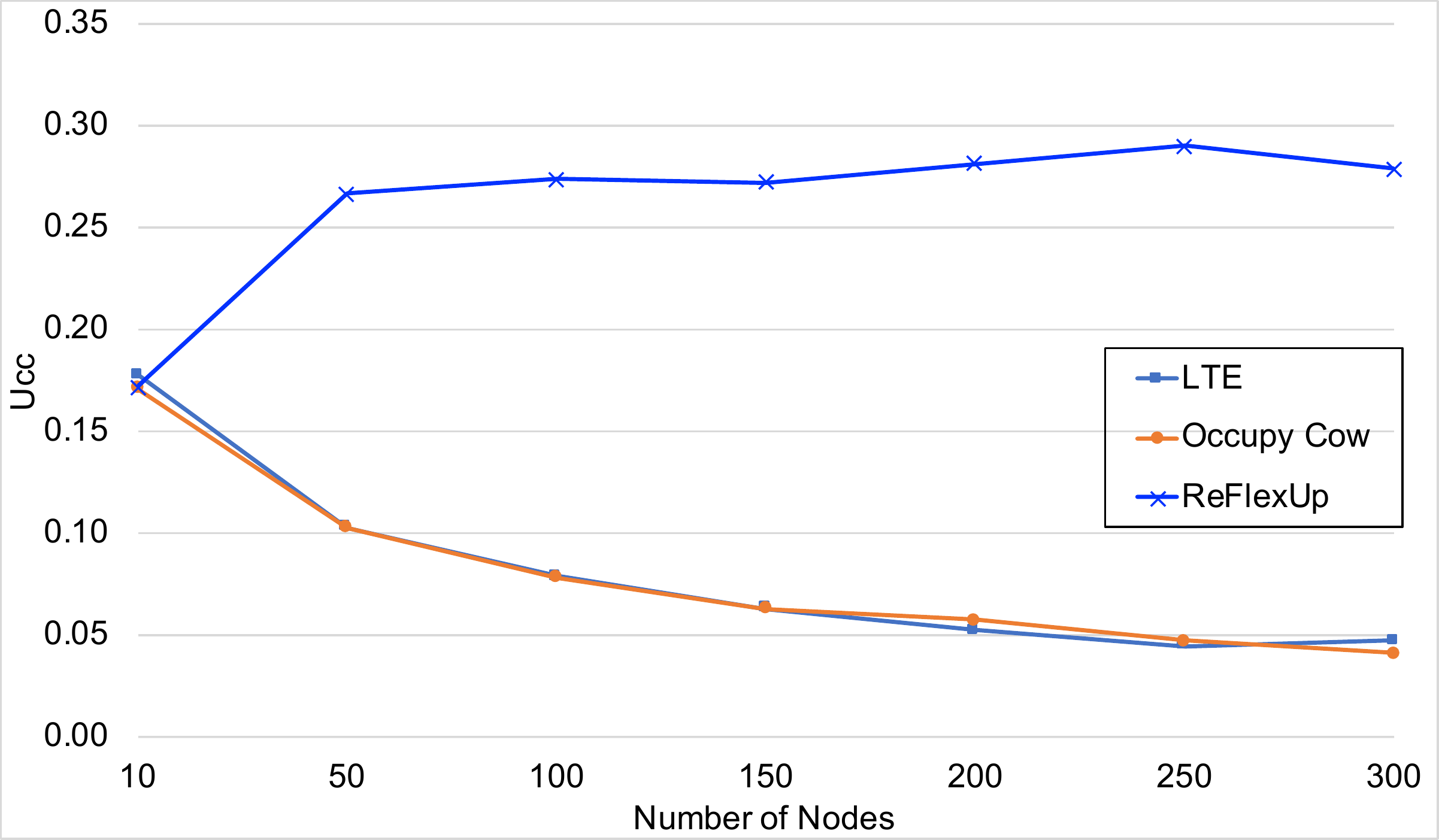}
\caption{$U_{cc}$ results of LTE, Occupy CoW, and ReFlexUp}
\label{FigSimuUcc}
\end{figure}

In Fig. \ref{FigSimuUcc}, we can see ReFlexUp outperforms the other two protocols while Occupy CoW and LTE have similar performance. The tendency of the $U_{cc}$ curve in Fig. \ref{FigSimuUcc} is similar to Fig. \ref{FigUccResult2} as ReFlexUp uses the optimal values of $T_{cm}$ and $T_{cp}$ to maximize the $U_{cc}$ value. 

\begin{figure}[!t]
\centering
\includegraphics[width=3.01in]{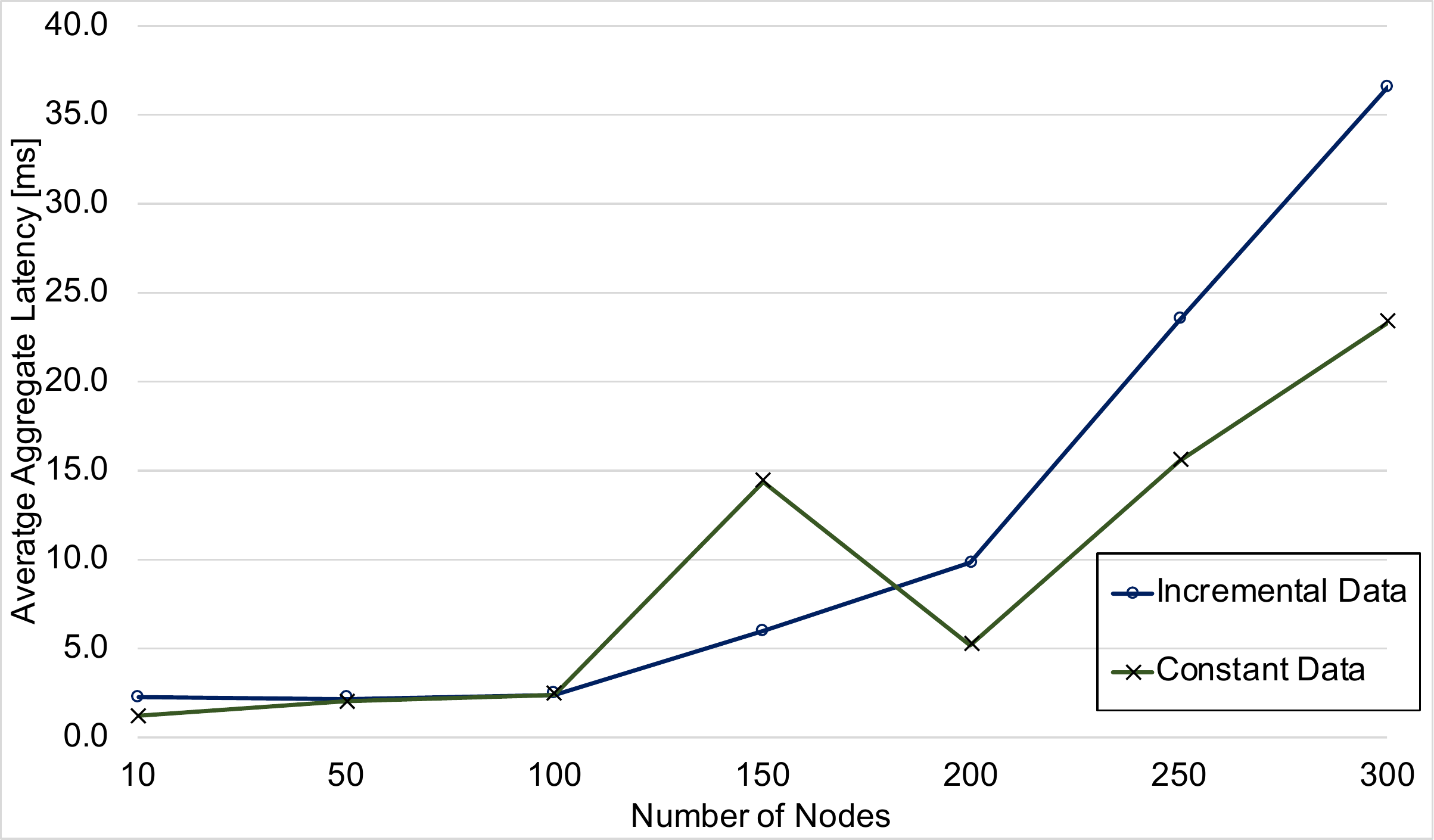}
\caption{Average latency results of ReFlexUp with incremental and constant data amounts on sensor nodes}
\label{FigSimuAvgLat}
\end{figure}

The average aggregate latency per UE node in shown in Fig. \ref{FigSimuAvgLat}, where two scenarios are tested: incremental data and constant data. In order to clearly see the how the latency guarantee is achieved with the different sizes of UE nodes, the data points on x-axis with a smaller time scale are plotted. The former scenario is to check the performance of a setup where the total amount of sensing data on each $v$ node linearly scales with the increasing network size ($N_{G}$); and the latter scenario is to check the performance of a setup where the total sensing data amount on each $v$ node does not scale with the increasing network size. Since the UE will receive the packets and transmit, in both scenarios there are constant number of sensors with either incremental or constant amount of data, the latency performance will behave differently. With the increasing UE nodes, the workload of the eNB nodes will increase, together with other factors such as the scheduling and buffering mechanism at UE nodes, and competition of the limited slots in a subframe. The turning points from N=100, N=150, to N=200 happen because of the increasing conflicts in the network where the radio resource scheduling at the RLC layer experiences conflicts from the UEs with the constant data scenario. At these data points, compared to the incremental data scenario, such conflicts do not appear because of the generally lighter data load.

\begin{figure}[!t]
\centering
\includegraphics[width=3.30in]{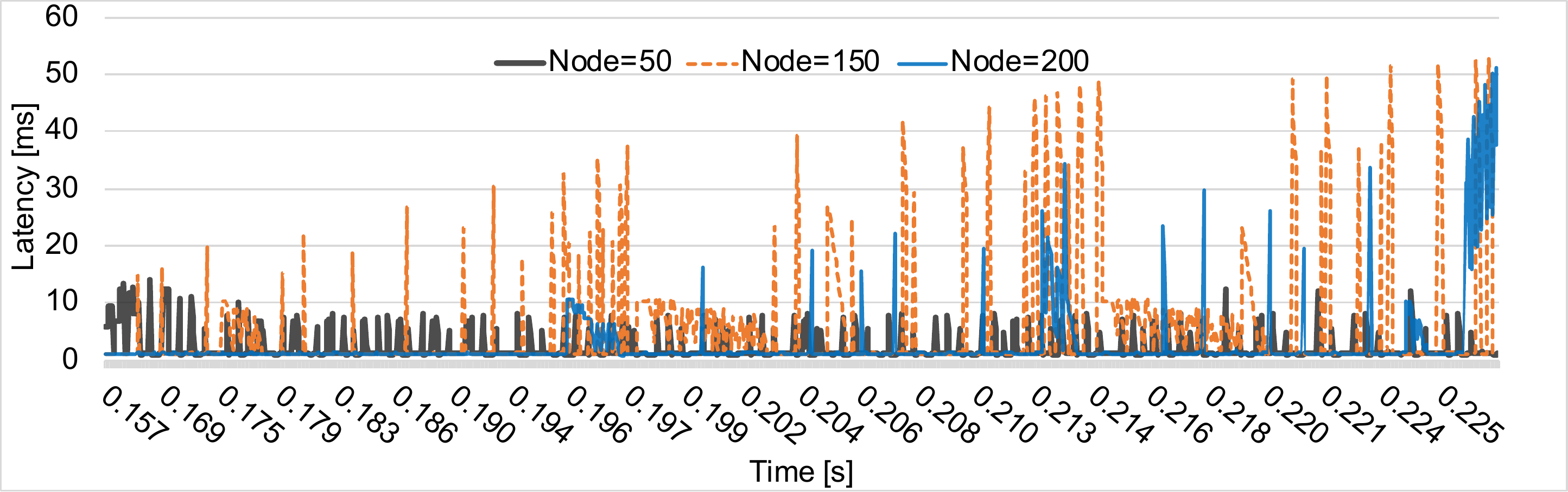}
\caption{Actual latency of data packet transmission over time}
\label{FigSimuLatActual}
\end{figure}

We can see the average latency is only affected by a small part of the transmissions if we check the specific per-packet latency at the Packet Data Convergence Protocol (PDCP) layer shown in Fig. \ref{FigSimuLatActual}, where, in fact when $N_{G}=50$, the latency performance is fairly good and there are only 1.48\% packets exceeding the 10 ms latency; when $N_{G}=150$, 10.1\% packets have over 10 ms latency; and when $N_{G}=200$, 22.95\% packets have over 10 ms latency. Although this phenomenon is expected and the overall goal of the $T_{cm}$ can successfully be maintained as required even if a challenging channel condition with modest temporal SNR values is adopted, it may be further optimized through, for example, the appropriate setup of $n_s$ and $n_v$, an enhanced channel condition, a dense deployment of eNBs, an offloading mechanism among eNBs, or in general an improved control over radio resource across the network. 

\section{Conclusion}
5G URLLC and MEC will significantly enhance the capabilities of wireless communication and computation which can enable numerous mission- and time-critical industrial applications for industrial automation applications. However, for a typical FDDS application with a real-world industrial process model in terms of the data flows and data processing models, we can see a URLLC-based system alone does not necessarily provide an efficient solution. A new 5G MEC-based architecture needs to be studied to ensure a successful URLLC-based solution to a broad range of FDDS-like industrial automation applications in the cellular IoT environment. The proposed ReFlexUp protocol in a CEC loop demonstrates the effectiveness of the proposed CEC efficiency measure although other non-greedy heuristic options may be available. As ReFlexUp is not deeply coupled with the lower-layer features, it is able to be extended to work with underlying standards-based networks, e.g., networks with various OSI layer 1 and layer 2 protocols. In addition, the analysis and modeling of the CEC loop can help address an IoT strategy for deploying future IIoT and Industry 4.0 applications based on 5G and MEC, including the important AI-driven predictive maintenance.

Within the problem domain arising from this work, on the one hand, the queuing and buffering effects on the relay nodes and the VPs at the edge and an extended model with additional parameters for the proposed CEC efficiency will be studied. On the other hand, the implementation of the 5G nodes and their planning (e.g., a possible dense deployment of mmWave eNBs) need to be fine-tuned for an application-specific field deployment in an industrial setting.


%

\ifCLASSOPTIONcaptionsoff
  \newpage
\fi



\bibliographystyle{IEEEtran}
\bibliography{IEEEabrv,./MyCollection}
%

%
%

%

%
%
\begin{IEEEbiographynophoto}
{Peng Hu}
received the Ph.D. degree in Electrical Engineering from Queen’s University, Canada. He is currently a Research Officer at the National Research Council of Canada. He serves as an associate editor of the Canadian Journal of Electrical and Computer Engineering. He has served as a member on the IEEE Sensors Standards committee and on the organizing and technical boards/committees of industry consortia and international conferences including AllSeen Alliance, DASH7, IEEE PIMRC’17, and IEEE AINA’15.

His current research interests include the industrial Internet of Things, AI-based edge computing, and low-power wireless networks.
\end{IEEEbiographynophoto}

\begin{IEEEbiographynophoto}
{Jinhuan Zhang}
received the Ph.D. degree in communication and information systems from the Wuhan University of Technology in 2009. Since 2009, she has been an Assistant Professor with the School of Information Science and Engineering, Central South University. Her research interest includes wireless sensor networks, resource management, and IoT.
\end{IEEEbiographynophoto}



\end{document}